\documentclass[fleqn,usenatbib]{mnras}

\usepackage{newtxtext,newtxmath}
\usepackage[normalem]{ulem}
\usepackage[T1]{fontenc}
\usepackage{ae,aecompl}
\usepackage{enumitem}

\usepackage{graphicx}	
\usepackage{amsmath}	
\usepackage{amssymb}	

\newcommand{\hi}{H\textsc{i}\ }
\newcommand{\hinospace}{\textrm{H\textsc{i}}}
\newcommand{\size}[2]{{\fontsize{#1}{0}\selectfont#2}}

\newcommand{\fastica}{\size{7.5}{FASTICA}}
\newcommand{\multidark}{\textsc{MultiDark}}
\newcommand{\deltadiff}{\delta\hspace{-0.3mm}}
\newcommand{\diff}{\text{d}\hspace{-0.25mm}}

\title[Multipole expansion for HI intensity mapping experiments]{Multipole expansion for HI intensity mapping experiments: simulations and modelling}

\author[S. Cunnington et al.]
{Steven Cunnington$^{1,2}$\thanks{E-mail: s.cunnington@qmul.ac.uk}, Alkistis Pourtsidou$^{1,3}$, Paula S. Soares$^{1}$,
Chris Blake$^{4}$
\newauthor
and David Bacon$^{2}$
\\
$^{1}$School of Physics and Astronomy, Queen Mary University of London, Mile End Road, London E1 4NS, UK\\
$^{2}$Institute of Cosmology \& Gravitation, University of Portsmouth, Dennis Sciama Building, Burnaby Road, Portsmouth PO1 3FX, UK\\
$^{3}$Department of Physics \& Astronomy, University of the Western Cape, Cape Town 7535, South Africa\\
$^{4}$Centre for Astrophysics \& Supercomputing, Swinburne University of Technology, P.O. Box 218, Hawthorn, VIC 3122, Australia
}

\date{Accepted XXX. Received YYY; in original form ZZZ}

\pubyear{2019}

\begin{document}
\label{firstpage}
\pagerange{\pageref{firstpage}--\pageref{lastpage}}
\maketitle

\begin{abstract}
We present a framework and an open-source \texttt{python} toolkit to analyse the 2-point statistics of 3D fluctuations in the context of \hi intensity maps using the multipole expansion formalism. We include simulations of the cosmological \hi signal using $N$-body and log-normal methods, foregrounds and their removal, as well as instrumental effects. Using these simulations and analytical modelling, we investigate the impact of foreground cleaning and the instrumental beam on the power spectrum multipoles as well as on the Fourier space clustering wedges. We find that both the instrumental beam and the foreground removal can produce a quadrupole (and a hexadecapole) signal, and demonstrate the importance of controlling and accurately modelling these effects for precision radio cosmology. We conclude that these effects can be modelled with reasonable accuracy using our multipole expansion technique.
We also perform an MCMC analysis to showcase the effect of foreground cleaning on the estimation of the \hi abundance and bias parameters. The accompanying \texttt{python} toolkit is available at \url{https://github.com/IntensityTools/MultipoleExpansion},
and includes an interactive suite of examples to aid new users. 
\end{abstract}



\section{Introduction}

A large portion of our understanding of the Universe comes from probing large scale cosmic structure using galaxy redshift surveys. Since galaxies trace the underlying dark matter density field, we can study the cosmic web's structure and evolution by mapping the position of the galaxies on the sky using angular coordinates and inferring a distance from their redshift. This approach has provided some excellent constraints on cosmological parameters (see e.g. \citet{Percival:2001hw, 2011MNRAS.418.1707B, Samushia:2013yga, Satpathy:2016tct, 2017MNRAS.470.2617A, 2018PhRvD..98d3526A}).

Obtaining accurate redshifts for resolved galaxies through spectroscopic analysis is time consuming and expensive. Consequently spectroscopic surveys are usually incomplete, insufficiently dense  samples, especially at high redshifts where they can be shot noise dominated. 
Another method is to use galaxies from imaging surveys where photometric redshifts can be inferred based on the amount of signal received through each of the telescopes broad colour bands. However, the error on these photometric redshifts can often be large and prone to systematic errors and therefore what is gained in increased sample size, is paid for with an increase in redshift-based distance uncertainty.

A promising alternative to galaxy redshift surveys comes from 21cm intensity mapping \citep{Battye2004,Chang:2007xk,Seo_2010,Pritchard:2011xb}. Neutral hydrogen (\hinospace) resides in many galaxies and spontaneously emits radiation from its single electron's ground state hyperfine transition. This radiation is emitted with an energy of $5.87 \, \mu$eV and hence has a rest wavelength of 21cm (equivalently, a 1420 MHz rest frequency). By detecting this signal with radio telescopes (single dishes or interferometers) we can effectively map large scale structure since \hi should be a reliable tracer of the underlying dark matter density \citep{Masui:2012zc}. Intensity mapping works by detecting the combined, unresolved 21cm emission from numerous galaxies binning them into low angular resolution maps. When using single dishes, the radio telescope beam (which modulates the survey's effective angular resolution) can be around or above the degree scale and therefore small angular scale information is lost. However, most of the scales of interest for probing large scale structure are of sufficient size that intensity mapping can still be used. Furthermore, the broad telescope beam means fewer pointings are needed to cover the target area of sky thus allowing large volumes of sky to be mapped very quickly.

A large challenge to overcome for \hi intensity mapping experiments lies in understanding and controlling the various instrumental and systematic effects, of which 21cm foregrounds is particularly demanding. In this context, foregrounds refer to galactic and extragalactic radio signals present in a similar frequency range to the \hi signals we are aiming to detect. These can be several orders of magnitude larger than the cosmological signal and thus their removal is required. The fact that foregrounds are often continuum signals and have smooth frequency coherence along the line-of-sight (LoS), provides a feature to distinguish them from the cosmological signal that is highly oscillatory with frequency. Previous work has provided very encouraging results suggesting foregrounds can be cleaned. However, foreground removal techniques invariably cause some unwanted consequences to the cosmological signal, e.g. by removing large scale power \citep{Wolz:2013wna,Alonso:2014dhk,Witzemann:2018cdx,Cunnington:2019lvb,Asorey:2020mxs}. Understanding the effects of foreground removal on our ability to use \hi intensity mapping observations for precision cosmology is therefore important and an active area of research.

The process of obtaining angular coordinates in large scale structure surveys is well understood and fairly straightforward. Obtaining a reliable radial distance based on redshift is more complicated. Even assuming an accurate redshift can be measured using spectroscopy, a well constrained distance-redshift relation is still required to obtain the third coordinate for the tracer data. Furthermore, if relying on redshifts, consideration must be given to the inherent peculiar velocity of the galaxy caused by local density perturbations. There are two contributions to the observed redshift $z_\text{obs}$ from both the cosmological Hubble flow $r\left(z_{\mathrm{cos}}\right)=\int_{0}^{z_\text{cos}} c\, \diff z / H(z)$ and the peculiar velocities such that
\begin{equation}\label{PecVelocityEq}
    1+z_\text{obs}=\left(1+z_\text{cos}\right)\left(1-\frac{v^\text{p}_\parallel(\vec{r})}{c}\right)^{-1}\, ,
\end{equation}
where the peculiar velocity $v_\parallel^\text{p}$ introduces more scatter in the measurement of objects closer to us. Since these peculiar velocities are correlated to density perturbations, any attempted measurement of a density field using redshift will therefore be distorted. The resulting distortions to the density correlations are known in the literature as Redshift Space Distortions (RSD).

The impact RSD have on the density field measured in redshift space is that on large scales, objects tend to fall in to high density regions which squashes the density field and the clustering amplitude becomes stronger along the line of sight (LoS) -- this is known as the \textit{Kaiser} effect \citep{Kaiser:1987qv}. RSD effects are also apparent in the non-linear regime on small scales.
There, objects are virialized, the density field becomes stretched and the clustering amplitude becomes smaller along the LoS -- this is called the Finger-of-God (FoG) effect \citep{Jackson:2008yv}.
While at first sight RSD might seem like a problem, they turn out to be extremely useful for measuring the logarithmic growth rate of structure $f$ \citep{Hamilton:1997zq}, which strongly depends on cosmology and gravity \citep{Guzzo:2008ac}.
For \hi intensity mapping, they can also be used to break the degeneracy between the \hi bias, $b_{\hinospace}$, and mean \hi abundance, $\Omega_{\hinospace}$ \citep{Masui:2012zc, Pourtsidou:2016dzn}.

The multipole expansion method represents a useful way to compress the data from the clustering statistics with respect to the local LoS. All the cosmological information is encoded in the first three even multipole moments in linear theory and thus represents a convenient way to analyse cosmological data. Furthermore, extensive work has been done to develop techniques that overcome complications induced by curved sky effects to allow multipole analysis on large-sky surveys (e.g. \citet{Bianchi:2015oia,Beutler:2016arn,Castorina:2017inr,Blake:2018tou}). Using clustering statistics as a function of the angle from the LoS has also been suggested to attempt to suppress survey systematics \citep{Reid:2014iaa,Hand:2017irw} or to avoid foreground contaminated regions in 21cm Epoch of Reionization studies \citep{Raut:2017zhh}.

In this work, we aim to investigate the prospects for probing the anisotropic \hi clustering using the expansion of the power spectrum into multipoles. While the use of power spectrum multipoles is standard practice in the data analysis of large scale structure surveys, the effect of 21cm foregrounds on the \hi monopole, quadrupole, and hexadecapole has not been investigated. We are particularly interested in the modelling of the signal and the effect of the instrumental beam and foreground removal on the multipoles. For this purpose we will extend upon the work in \citet{Blake:2019ddd} (hereafter B19), which looked at modelling the power spectrum for galaxy and \hi intensity map data including observational effects. Here we focus on the \hi auto-power spectrum, and we include more sophisticated simulations of the \hi signal as well as foregrounds and their removal. We aim to provide analytical phenomenological models to describe foreground removal and other observational effects, and then test these with measurements from our simulations.
The outline of this paper is as follows: in Section~\ref{ModellingSec} we introduce a theoretical model for the \hi multipoles, which aims to 
be able to emulate the expected impact from the telescope beam and foreground contamination and removal. In Section~\ref{SimulationsSec} we present our approach for simulating cosmological \hi signal data along with relevant observational effects and 21cm foregrounds. We then discuss the method we use for removing these foregrounds. In Section~\ref{ResultsSec} we present our results from this analysis that demonstrate how observational effects impact the \hi power spectrum multipoles, and how important it is to take them into account in parameter estimation studies. To illustrate this point, we perform an MCMC analysis to estimate the \hi parameters in the presence of foregrounds. We summarise and conclude in Section~\ref{ConclusionSec}.

\section{Modelling Observational Effects}\label{ModellingSec}

\subsection{Power Spectrum Multipoles}

To account for RSD we use the anisotropic power spectrum which is explicitly dependent on the direction of the wave vector $\vec{k}$ relative to the LoS. This can be written as
\begin{equation}\label{RSDPk}
	P_\hinospace(\vec{k}) \equiv P_\hinospace(k, \mu) = \overline{T}_\hinospace^2\left[\frac{b_\hinospace^2 \left(1+\beta \mu^{2}\right)^{2} P_\text{M}(k)}{1+\left(k \mu \sigma_\text{v} / H_{0}\right)^{2}} + P_\text{SN}\right] \, .
\end{equation}
The $\mu$-dependent terms account for the anisotropic effect of RSD, where $\mu$ is defined as the cosine of the angle $\theta$ between the LoS and $\vec{k}$, i.e. $\mu \equiv \cos\theta$. This means that modes perpendicular (parallel) to the LoS have $\mu=0$ ($\mu=1$). Equation \eqref{RSDPk} also depends on the bias of the \hi tracer $b_\hinospace$, the mean \hi temperature $\overline{T}_\hinospace$ and $\beta = f/b_\hinospace$, where $f$ is the linear growth rate of structure which can be approximated by $f \simeq \Omega_\text{M}(z)^\gamma$ where $\gamma$ is the growth rate index \citep{Linder:2005in}. We note that $\Omega_{\mathrm{M}}(z)=H_{0}^{2} \Omega_{\mathrm{M},0}(1+z)^{3} / H(z)^{2}$ and $\gamma \simeq 0.55$ for $\Lambda$CDM. The factor on the denominator accounts for the FoG effect and $\sigma_\text{v}$ is the velocity dispersion of the tracers. For a full derivation of the above we refer the reader to the review in \citet{Hamilton:1997zq}, while examples of studies of the FoG effect for \hi intensity mapping are \citet{Sarkar:2018gcb,Sarkar:2019nak}. The term $P_\text{SN}$, is the Poisson shot noise, which appears because of the fact that a finite number of galaxies is used to probe a continuous density field. This is a scale invariant term and in optical galaxy surveys it is simply the inverse of the galaxy density in a redshift bin, i.e. $P_\text{SN} = 1/\overline{n}_\text{g}$. In intensity mapping the shot noise effect is expected to be negligible, especially with respect to the instrumental noise contribution, since every  galaxy with \hi content contributes to the total intensity map signal. 

The anisotropic power spectrum $P_\hinospace(k, \mu)$ can be expanded in Legendre polynomials as
\begin{equation}\label{LegExpansion1}
	P_\hinospace(k, \mu)=\sum_{\ell} P_{\ell}(k) \mathcal{L}_{\ell}(\mu) \, ,
\end{equation}
where $\mathcal{L}_{\ell}(\mu)$ is the $\ell^\text{th}$ Legendre polynomial. Given that the Legendre polynomials are orthogonal over $[-1,1]$, we have the identity
\begin{equation}
	\int_{-1}^{1} \mathcal{L}_{\ell}(\mu) \mathcal{L}_{m}(\mu) \diff \mu=\frac{2}{2\ell+1} \delta_{\ell\hspace{-0.2mm} m} \, ,
\end{equation}
where $\delta_{\ell\hspace{-0.2mm} m}$ is the Kronecker delta. By multiplying both sides of \eqref{LegExpansion1} by an orthogonal Legendre polynomial, integrating and then rearranging we can derive a general expression for the \emph{power spectrum multipoles} given by
\begin{equation}\label{LegExpansion2}
	P_{\ell}(k)=\frac{2 \ell+1}{2} \int_{-1}^{1} \diff \mu P_\hinospace(k, \mu) \mathcal{L}_{\ell}(\mu).
\end{equation}
In linear theory, the only non-zero power spectrum multipoles are given by $\ell = 0,2,4$ (monopole $P_0$, quadrupole $P_2$, and hexadecapole $P_4$), and even when non-linearities are taken into account these multipoles contain most of the cosmological information (see e.g. \citet{Taruya:2011tz}). Therefore, the Legendre polynomials we need are given by
\begin{equation}\label{LegPolys}
	\mathcal{L}_{0} = 1, \quad \mathcal{L}_{2} = \frac{3\mu^2 -1}{2}, \quad \mathcal{L}_{4} = \frac{35\mu^4 -30\mu^2 + 3}{8} \, .
\end{equation}
In this work, we investigate how the power spectrum multipole measurements are influenced by the main observational effects relevant to single dish intensity mapping surveys (with instruments like MeerKAT \citep{Pourtsidou:2017era} and SKA-MID \citep{Santos:2015bsa}) from foreground contamination and the telescope beam. The beam effect can be modelled as a convolution of the density field and the Fourier transform of this smoothing term is given as \citep{Villaescusa-Navarro:2016kbz}
\begin{equation}\label{BeamDampEq}
	\tilde{B}_{\mathrm{beam}}(k,\mu) = \exp\left(\frac{-k^2 R_\text{beam}^2(1-\mu^2)}{2}\right) \, ,
\end{equation}
where $R_\text{beam}$ is the scale of the beam at the effective central redshift of the survey. This is defined as $R_\text{beam} = \sigma_\theta\chi(z_\text{eff})$, where $\sigma_\theta =  \theta_\text{FWHM} / ( 2\sqrt{2\ln(2)} )$ and $\theta_\text{FWHM}$ is the full-width-half-maximum of the beam in radians. We also need to take into account the instrumental (thermal) noise from the telescope, $P_\text{N}$, which will be described in detail later on.

As shown in B19, further consideration could be given to the damping from the binning of the \hi data into angular pixels and frequency channels. However, in this work we will be using a flat-sky, Cartesian data cube (we discuss the reasons for this in Section \ref{SimulationsSec}). Therefore we will simply correct for the aliasing effect following \citet{Jing:2004fq} -- this comes from sampling effects when using a \emph{mass assignment function} to assign the particle distribution onto grids. Since our resulting Fast Fourier Transform (FFT) grid will have the same dimensions as our intensity map data this should sufficiently encapsulate all effects from discretization. Were there to be some re-gridding, necessary in cases where large-sky lightcone data are transformed into a Fourier cuboid, then further consideration would be needed to properly deal with angular pixelization (see B19 for details). For our purpose the Fourier transform of the mass assignment function is given by
\begin{equation}
	\tilde{W}_{\mathrm{grid}}(\vec{k}) = \left[ \frac{\sin \left(k_\text{x} H_\text{x}/2\right)}{\left(k_\text{x} H_\text{x} / 2 \right)} \frac{\sin \left(k_\text{y} H_\text{y}/2\right)}{\left(k_\text{y} H_\text{y}/2\right)} \frac{\sin \left(k_\text{z} H_\text{z}/2\right)}{\left(k_\text{z} H_\text{z}/2\right)} \right]^p \, ,
\end{equation}
and it is sufficient to account for the damping from this discretization; consequently, our resulting measurements will be divided through by $\tilde{W}_{\mathrm{grid}}^2$. Here $H_i = L_i/N_i$ is the grid spacing where $L_i$ define the comoving size of our Fourier cuboid and $N_i$ the number of pixels in each dimension. The choice of $p$ relates to the mass-assignment method chosen (see \citet{Jing:2004fq} for details). For this work we use the Nearest Grid Point (NGP) assignment with $p=1$.

The damping factor from the beam can be applied to the power spectrum and using the Legendre polynomials from equation \eqref{LegPolys} we expand the anisotropic power spectrum (equation \eqref{LegExpansion2}) giving the formulae for the monopole ($P_0$), quadrupole ($P_2$), and hexadecapole ($P_4$) as:
\begin{equation}\label{ObsMultipole0}
\begin{split}
    	P_{0}(k)=\frac{1}{2} \overline{T}_\hinospace^2\, \Biggl[ \int_{-1}^{1} \diff \mu  \ & \frac{b_\hinospace^2 \left(1+\beta\mu^2\right)^2P_\text{M}(k)\,\mathcal{L}_0 \, \tilde{B}^2_{\mathrm{beam}}}{1+\left(k \mu \sigma_\text{v} / H_{0}\right)^{2}} \\
    	\ & + \int_{-1}^{1} \diff \mu\, P_\text{SN}\,\mathcal{L}_0 \, \tilde{B}^2_{\mathrm{beam}} \Biggl]\, ,
\end{split}
\end{equation}
\begin{equation}\label{ObsMultipole2}
\begin{split}
    	P_{2}(k)=\frac{5}{2} \overline{T}_\hinospace^2\, \Biggl[ \int_{-1}^{1} \diff \mu  \ & \frac{b_\hinospace^2 \left(1+\beta\mu^2\right)^2P_\text{M}(k)\,\mathcal{L}_2 \, \tilde{B}^2_{\mathrm{beam}}}{1+\left(k \mu \sigma_\text{v} / H_{0}\right)^{2}} \\
    	\ & + \int_{-1}^{1} \diff \mu\, P_\text{SN}\,\mathcal{L}_2 \, \tilde{B}^2_{\mathrm{beam}} \Biggl]\, ,
\end{split}
\end{equation}
\begin{equation}\label{ObsMultipole4}
\begin{split}
    	P_{4}(k)=\frac{9}{2} \overline{T}_\hinospace^2\, \Biggl[ \int_{-1}^{1} \diff \mu  \ & \frac{b_\hinospace^2 \left(1+\beta\mu^2\right)^2P_\text{M}(k)\,\mathcal{L}_4 \, \tilde{B}^2_{\mathrm{beam}}}{1+\left(k \mu \sigma_\text{v} / H_{0}\right)^{2}} \\
    	\ & + \int_{-1}^{1} \diff \mu\, P_\text{SN}\,\mathcal{L}_4 \, \tilde{B}^2_{\mathrm{beam}} \Biggl]\, .
\end{split}
\end{equation}
Note that in the model, the shot noise term $P_\text{SN}$ from equation \eqref{RSDPk} is also damped with the telescope beam and is weighted by the \hi content of each galaxy ($\overline{T}_\hinospace \propto \Omega_\hinospace$). We discuss the shot noise further in Section~\ref{SimulationsSec}, along with our method for treatment of the instrumental noise $P_\text{N}$ in our simulations.

\subsubsection{Clustering Wedges}\label{WedgesSec}

An alternative method for compressing the variation of the power spectrum with respect to the LoS is to measure the power spectrum (or correlation function) in a wedge-shaped region \citep{Kazin:2011xt, Grieb:2016uuo, Sanchez:2016sas, Hand:2017irw}, for example to break degeneracies \citep{Jennings:2015lea}. The power spectrum wedge in the region $\mu_1 \leq \mu \leq \mu_2$ is given by
\begin{equation}
P^{\rm wedge}(k) = \frac{1}{\mu_2-\mu_1}\int_{\mu_1}^{\mu_2} \diff\mu\,P_\hinospace(k,\mu) \, . 
\label{eq:Pwedge}
\end{equation}
Using these clustering wedges as a means for avoiding foreground contaminated regions present in 21cm interferometer observations of the Epoch of Reionization has been investigated in \citet{Raut:2017zhh}. We will build on this concept and analyse our simulations in different clustering wedges with the aim of understanding how foregrounds are having an impact. 

\subsection{Modelling Signal Loss from 21cm Foregrounds}\label{MultipoleFGimpactSec}

\begin{figure}
	\centering
  	\includegraphics[width=\columnwidth]{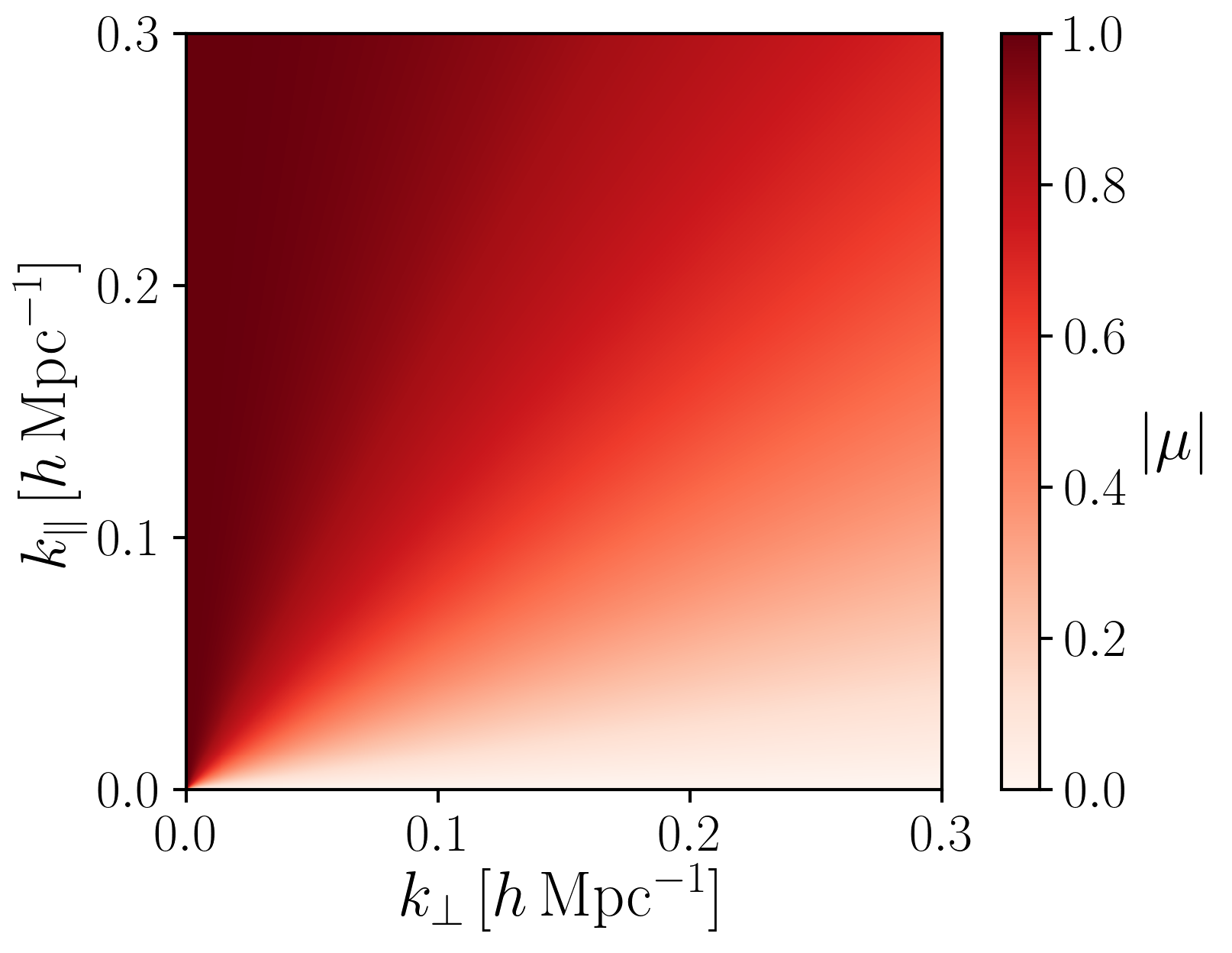}
    \caption{This plot demonstrates how $\mu$, the directional cosine of modes, changes depending on the contributions from modes parallel and perpendicular to the LoS. This is calculated from $\mu = \cos{\theta} = k_\parallel/k = k_\parallel / \sqrt{k_\parallel^2 + k_\perp^2}$.}
\label{muMap}
\end{figure}

\begin{figure*}
	\centering
  	\includegraphics[width=2.1\columnwidth]{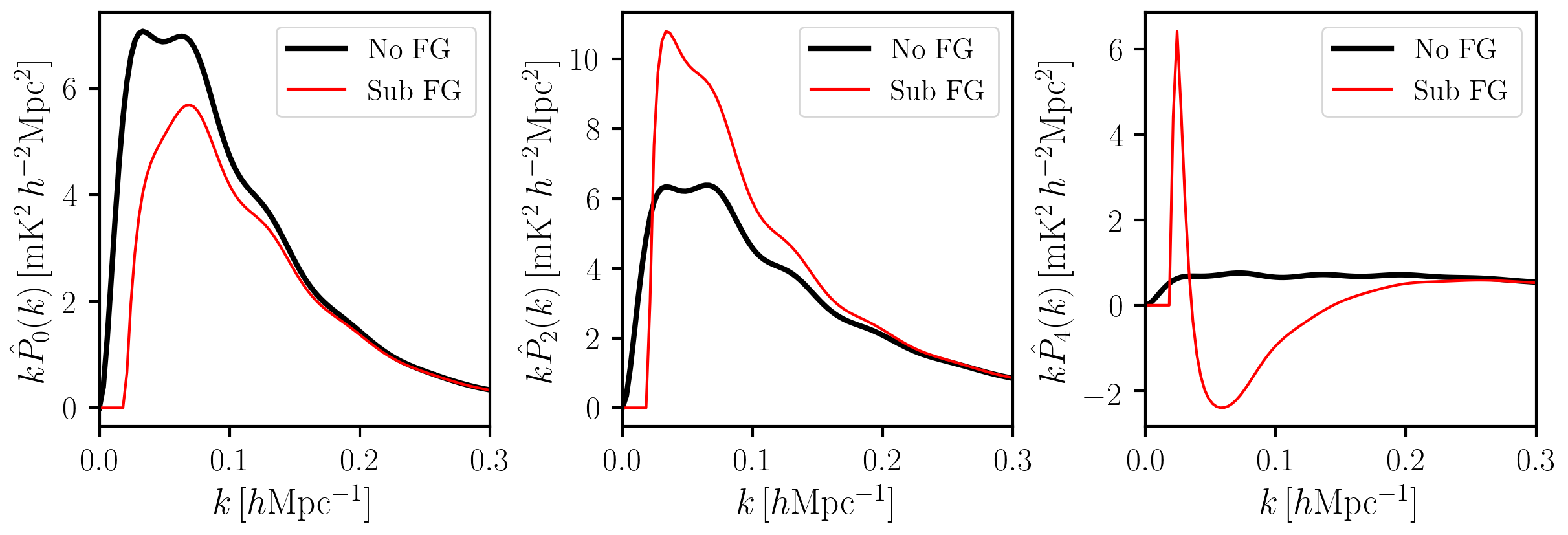}
    \caption{Theoretical power spectrum multipoles including removal of low-$\mu$ contributions to emulate a foreground clean as defined by equation~(\ref{ToyFGremovalEq}). The foreground subtracted cases are shown as thin red lines. These use a $\mu_\text{FG}$ cut-off defined by $\mu_\text{FG} = k_\parallel^\text{FG}/k$ with $k_\parallel^\text{FG} = 0.02\,h\text{Mpc}^{-1}$. We also show the input (true), foreground-free \hi signal for comparison (thick black lines) with $\mu_\text{FG}=0$. We have also employed a  $\theta_\text{FWHM}=0.44$ deg beam effect, which causes damping as outlined by equation~\eqref{BeamDampEq}.}
\label{toyFGmuremoval}
\end{figure*}

Since we expect foregrounds to have relatively smooth fluctuations along the LoS, a conventional, blind foreground clean will always remove cosmological power spectrum modes at small $k_\parallel$ below some $k^\text{FG}_\parallel$ cut-off, since these are the ones that will be indistinguishable from the foregrounds \citep{Wolz:2013wna,Shaw:2014khi,Alonso:2017dgh}. In Figure~\ref{muMap} we show how values of $\mu$ depend on the contributions from $k_\parallel$ and $k_\perp$, which are the modes parallel and perpendicular to the LoS respectively. Based on the above discussion, we can claim that foregrounds mostly affect low-$\mu$ modes, where $k \lesssim k^\text{FG}_\parallel$. Therefore, we can emulate the effect of their removal by limiting the $\mu$ parameter space we integrate over in equations~\eqref{ObsMultipole0}, \eqref{ObsMultipole2} and \eqref{ObsMultipole4} such that each multipole is given by
\begin{equation}\label{ToyFGremovalEq}
	\hat{P}_\ell(k)=(2 \ell+1) \int_{\mu=\mu_\text{FG}}^{\mu=1} \diff \mu\, P_\hinospace(k, \mu) \mathcal{L}_{\ell}(\mu)\tilde{B}^2_{\mathrm{beam}}\, .
\end{equation}
Note we use the notation $\hat{P}_\ell$ to emphasize that this is not the same as the conventional power spectrum multipoles but signifies our model power spectra where for foreground affected cases, $\hat{P}_\ell$ excludes the power from all modes with $\mu < \mu_\text{FG}$, with
\begin{equation}\label{muFGeq}
	\mu_\text{FG} = k_\parallel^\text{FG}/k
\end{equation}
and $k_\parallel^\text{FG}$ is the parallel wave-vector cutoff below which the foreground removal effects are expected to be more severe. If no foregrounds are present or in the idealised case where they are assumed to be perfectly cleaned without any signal loss, we have $\mu_\text{FG} = 0$ and recover the standard multipole expansion equation. 

We demonstrate the results from this approach in Figure~\ref{toyFGmuremoval}. The theoretical multipoles $\hat{P}_\ell(k)$ are produced from an underlying non-linear matter power spectrum generated using \texttt{Nbodykit}\footnote{\href{https://nbodykit.readthedocs.io/en/latest/}{https://nbodykit.readthedocs.io}} \citep{Hand:2017pqn} with Astropy\footnote{\href{http://www.astropy.org}{www.astropy.org}} \citep{Robitaille:2013mpa,Price-Whelan:2018hus}, the CLASS Boltzmann solver \citep{Lesgourgues:2011re,Blas:2011rf}, and the HaloFit prescription \citep{Takahashi:2012em}. For these models we use a \textsc{Planck}15 cosmology \citep{Ade:2015xua}, with $\Omega_\text{M} = 0.307$, and Hubble parameter $h=0.678$. These results are computed at $z=0.8$, assume a $\sigma_\text{v} = 400\,\text{km}\,\text{s}^{-1}$ velocity dispersion and for the foreground effected cases (red lines) we use $k_\parallel^\text{FG} = 0.02\,h\,\text{Mpc}^{-1}$ consistent with previous work \citep{Shaw:2014khi}. For the \hi parameters we use $b_\hinospace = 1$ and $\overline{T}_\hinospace=0.127\,\text{mK}$. We discuss these parameters in more detail in Section~\ref{SimulationsSec} when we introduce the simulated data and also tune their values to maximise agreement with the simulated data in Section \ref{ResultsSec}.

The results from this toy model in Figure \ref{toyFGmuremoval} reveal some interesting features, which we also see in our simulation results (discussed in Section~\ref{ResultsSec}). While we expect that foreground cleaning damps power, when considering the different power spectrum multipoles the quadrupole ($P_2$) result shows that the measured large scale power weighted as a function of $\mu$ with foreground removal is actually enhanced. 
The hexadecapole ($P_4$) also shows some interesting features with the model predicting a change of sign for large parts of the signal. The monopole ($P_0$) result is as expected, with foreground removal damping power, especially at large scales (small $k$).

\begin{figure}
	\centering
  	\includegraphics[width=\columnwidth]{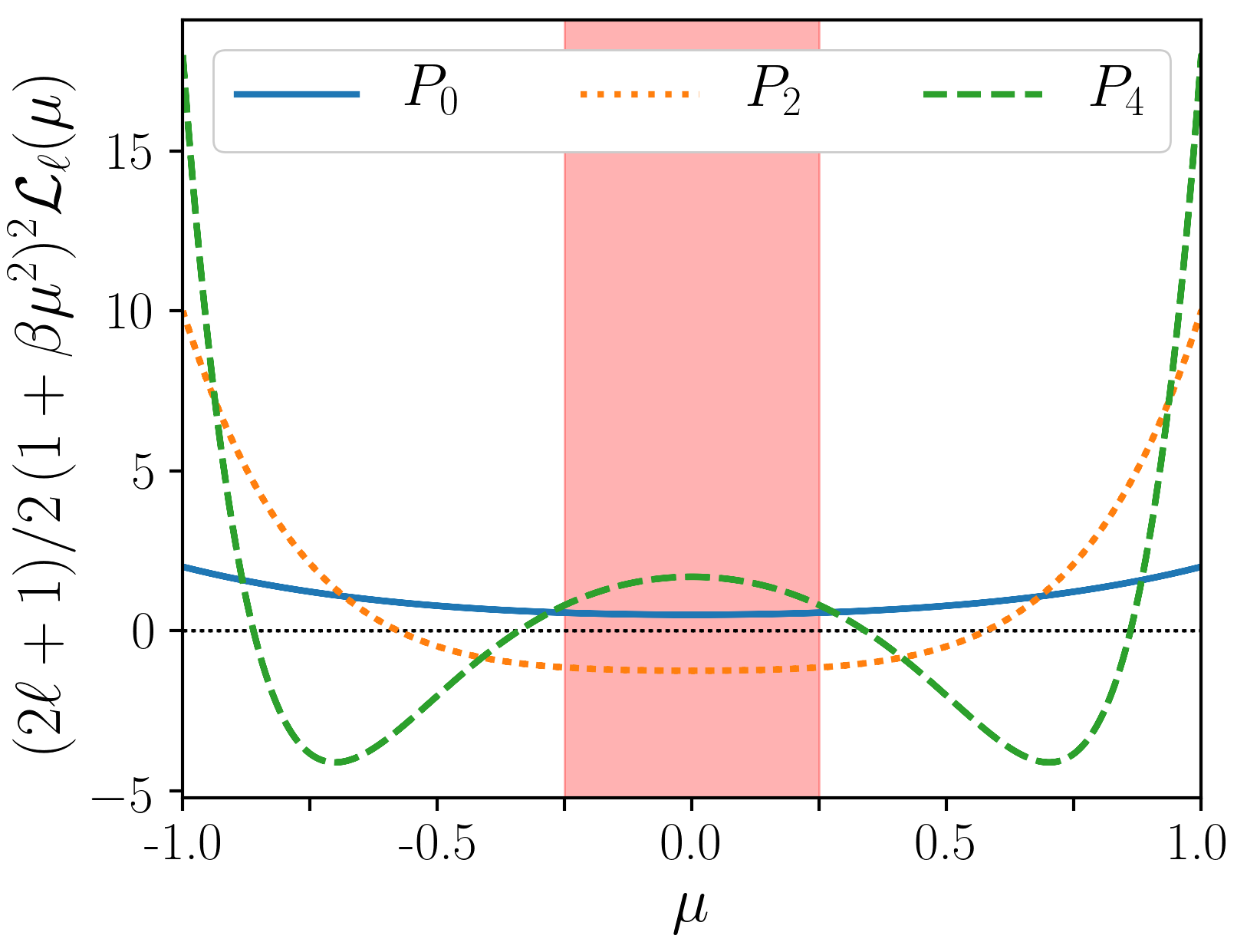}
    \caption{Expanded multipole functions as a function of $\mu$ for each $\ell=0,2,4$ (neglecting the FoG contribution). These functions are integrated over to form the multipoles in equations \eqref{ObsMultipole0}, \eqref{ObsMultipole2} and \eqref{ObsMultipole4}. The pink central shaded region has $|\mu|<0.25$ where large radial modes will dominate. Since this plot is for demonstrative purposes, we have used a dummy value of $\beta= 1$.}
\label{MultipoleIntegrands}
\end{figure}

These results can be understood by analysing how the expanded multipoles (equations \eqref{ObsMultipole0}, \eqref{ObsMultipole2} and \eqref{ObsMultipole4}) vary as a function of $\mu$. For this demonstration we will ignore the FoG factor. In Figure~\ref{MultipoleIntegrands} we have plotted the function $(2\ell + 1)/2\,(1+\beta\mu^2)^2\mathcal{L}_{\ell}(\mu)$ for varying $\mu$ for each multipole $\ell = 0,2,4$. This function is integrated over $\mu$ and therefore its behaviour can explain some of the results we are seeing in Figure~\ref{toyFGmuremoval}, since a foreground clean should have a similar effect to removing contributions to the multipoles from low-$\mu$ regions (e.g. $\mu<0.25$ shown as the pink shaded region). Doing this removes a lot of the negative contribution in the quadrupole which is why we see an enhanced signal. Similarly, this also removes positive contributions to the monopole, hence why we see an overall damping here and the hexadecapole has enough positive contributions removed for its negative contributions to dominate.

\subsection{Increasing Beam}\label{BigBeamModelSec}

\begin{figure*}
	\centering
  	\includegraphics[width=2.1\columnwidth]{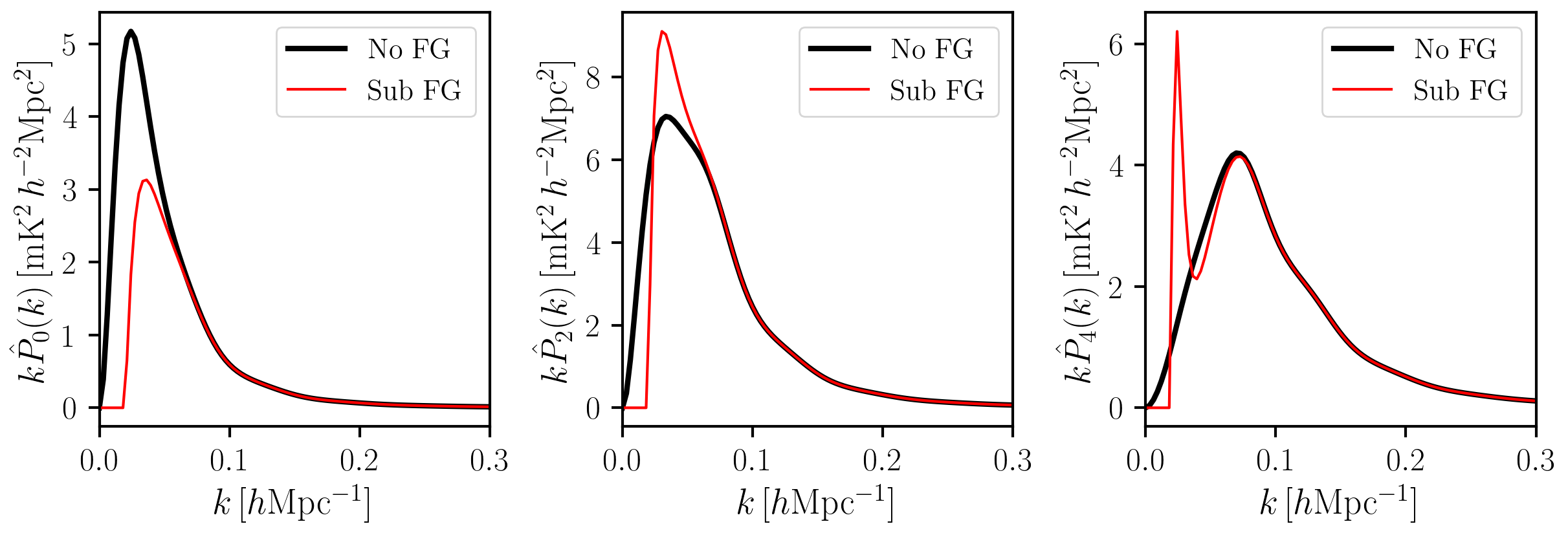}
    \caption{Same plot as Figure~\ref{toyFGmuremoval} but with an increased beam size of $\theta_\text{FWHM} = 2$ deg. We see more damping here at high-$k$ in comparison with Figure~\ref{toyFGmuremoval} as expected from equation \eqref{BeamDampEq}. The larger beam also lessens the effects from foregrounds in the quadrupole and hexadecapole in comparison to the smaller beam case of Figure~\ref{toyFGmuremoval}.}
\label{ToyFGbigBeam}
\end{figure*}

It is interesting to look at how the toy model forecasts presented in Figure~\ref{toyFGmuremoval} change when the size of the telescope beam is increased. For the results in Figure~\ref{toyFGmuremoval} we used a beam of $\theta_\text{FWHM}=0.44\,\text{deg}$. This corresponds to the size of a GBT-like beam \citep{Wolz:2015lwa}. Effective beam sizes are expected to be smaller than this for interferometers such as HIRAX \citep{Newburgh:2016mwi} and CHIME \citep{Bandura:2014gwa}, which can achieve a much better resolution than single dish experiments. However, beam sizes larger than $\theta_\text{FWHM}=0.44\,\text{deg}$ are expected on instruments such as SKA-MID \citep{Bacon:2018dui} or its pathfinder MeerKAT \citep{Santos:2017qgq}, which plan to operate in single-dish mode for intensity mapping surveys.

Figure~\ref{ToyFGbigBeam} demonstrates the same model as Figure~\ref{toyFGmuremoval} but with a larger beam of $\theta_\text{FWHM}=2\,\text{deg}$, and we can immediately see some differences.  Unsurprisingly, we see that the enhanced damping from the larger beam affects more mid-range values of $k$. That is because a larger beam effectively smooths out larger perpendicular modes, thus affecting smaller $k_\perp$. The difference between the foreground free and foreground contaminated cases is less intuitive. It appears that increasing the beam renders the difference between foreground free maps and foreground cleaned maps to be minimal when compared with the smaller beam of Figure~\ref{toyFGmuremoval}. There are still noticeable effects, but they are restricted to the small-$k$ region.

\begin{figure*}
	\centering
  	\includegraphics[width=1.6\columnwidth]{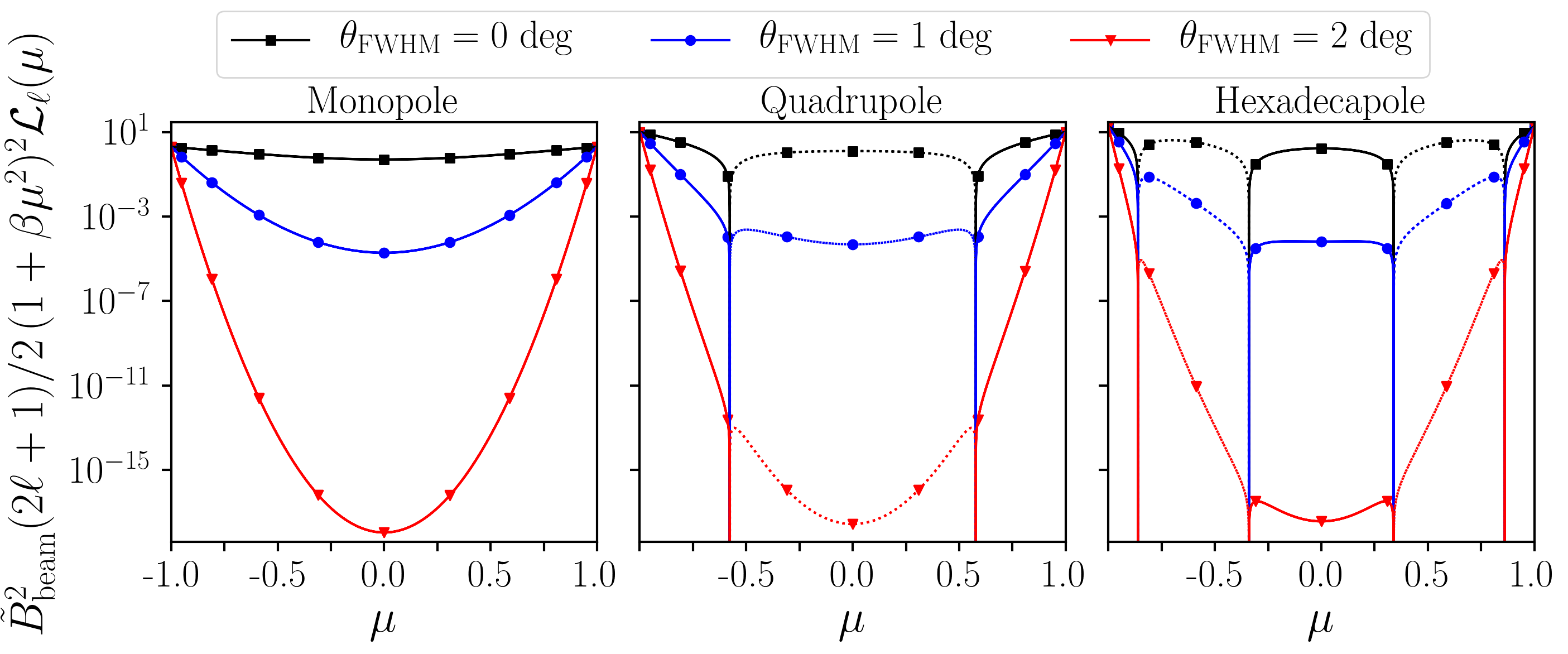}
    \caption{Effect of a varying beam size on the multipoles. Similarly to Figure~\ref{MultipoleIntegrands}, this shows the
    expanded multipole functions as a function of $\mu$ for each $\ell=0,2,4$, also including the effect of increasing the beam $\theta_\text{FWHM}$. Dotted lines represent negative values. These results are for a fixed value of $k=0.15\,h\text{Mpc}^{-1}$.}
\label{LIntegrandsSig}
\end{figure*}

To understand this we can again analyse the contribution to the multipoles from the function $(2\ell + 1)/2\,(1+\beta\mu^2)^2\mathcal{L}_{\ell}(\mu)$ as a function of $\mu$ as done in Figure~\ref{MultipoleIntegrands}. However, this time we show how damping from the beam (see equation \eqref{BeamDampEq}) modulated by the beam size $\theta_\text{FWHM}$, affects these functions. These results are shown in Figure~\ref{LIntegrandsSig} for a range of beam sizes. Since the beam damping term is dependent on $k$, we have chosen a fixed mid-range value ($k=0.15\,h\text{Mpc}^{-1}$) to demonstrate these effects. \footnote{As one would expect, we find that larger values of $k$ are affected more by the beam since the beam smooths small perpendicular scales, thus affecting large $k_\perp$ modes. Choosing a very small $k$-value for the results in Figure~\ref{LIntegrandsSig} would show little difference between each different $\theta_\text{FWHM}$ case.}

Figure~\ref{LIntegrandsSig} shows that a larger beam damps contributions across all $\mu$ values, but it has more of an effect at low-$|\mu|$. It is the modes with low-$|\mu|$ which are most affected by foregrounds and this is why we see apparent mitigation of foreground effects for intensity maps with large beams. It is simply because the beam is damping foreground contaminated modes anyway, rendering the foreground removal effects less dominant. For lower values of $k$, it is more likely that there will be smaller $k_\perp$ values which are less affected by the beam. Figure~\ref{muMap} shows that high values of $\mu$ exist mostly at these low-$k_\perp$ values where there is much less beam damping and this allows foreground effects to dominate. This is why we still see some foreground effects at low-$k$ in Figure~\ref{ToyFGbigBeam}.

\section{Simulations}\label{SimulationsSec}

This section explains how we generate our data for the simulated intensity maps along with the method for adding foregrounds and removing them with a \fastica\ reconstruction process. This data will be used to measure the expanded power spectrum multipoles with the results compared to the model outlined in Section~\ref{ModellingSec} and the forecasts presented therein.

For this work, we choose to work with flat-skies in Cartesian coordinates as opposed to curved-sky lightcone data. This avoids curved-sky complications such as wide-angle effects or more complex survey window functions \citep{Blake:2018tou}. The flat-sky choice means our maps can be constructed into Cartesian data-cubes with dimensions $[N_\text{x}\, ,\, N_\text{y}\, ,\, N_\text{z}]$. The total number of voxels (volume pixels) in the data cube is therefore given by $N_\text{vox} = N_\text{x}\,\times\, N_\text{y}\, \times\, N_\text{z}$. The data cube has co-moving physical dimensions $L_\text{x}, L_\text{y}$ and $L_\text{z}$, and therefore each voxel has a volume defined by $V_\text{vox} = L_\text{x} \times L_\text{y} \times L_\text{z} / N_\text{vox}$. The radial centre of this data cube lies at a comoving distance $\chi(z)$, where $z$ is the redshift of our simulated data. Note the distinction between the Cartesian coordinate z and the italicised $z$ that denotes redshift.

\subsection{Cosmological Signal}\label{CosmoSigSec}

Onto this grid we then bin galaxies which are drawn from a pre-simulated galaxy catalogue. Here we use the \textsc{MultiDark-Galaxies} data \citep{Knebe:2017eei} and the catalogue produced from the \textsc{SAGE} \citep{Croton:2016etl} semi-analytical model application. These galaxies were produced from the dark matter cosmological simulation \textsc{MultiDark-Planck} (MDPL2) \citep{Klypin:2014kpa}, which follows the evolution of 3840$^3$ particles in a cubical volume of 1$h^{-1}$Gpc$^3$ with mass resolution of 1.51$\,\times\, 10^9h^{-1}$M$_\odot$ per dark matter particle. The cosmology adopted for this simulation is based on \textsc{Planck}15 cosmological parameters \citep{Ade:2015xua}, with $\Omega_\text{M} = 0.307$, $\Omega_\text{b} = 0.048$, $\Omega_\Lambda = 0.693$, $\sigma_8 = 0.823$, $n_\text{s} = 0.96$ and Hubble parameter $h=0.678$. The catalogues
are split into 126 snapshots between redshifts $z= 17$ and $z=0$. In this work we want to utilise the lower redshift data in the post re-ionization Universe. In particular we use two snapshots at $z=0.82$ and at $z=2.03$. We obtained this publicly available data from the Skies \& Universes web page\footnote{\href{http://www.skiesanduniverses.org/page/page-3/page-22/}{www.skiesanduniverses.org}}.

From a redshift snapshot of this simulation we extract each of the galaxy coordinates (x, y, z) in $\text{Mpc}\,h^{-1}$ that define the galaxy's position $\vec{r}$. To simulate RSD we assume the LoS is along the z-dimension and use the plane parallel approximation to displace the galaxy positions to a new coordinate $z_\text{RSD}$ given by
\begin{equation}\label{PlaneParaEq}
    \text{z}_\text{RSD} = \text{z} + \frac{1+z}{H(z)}h\, v^\text{p}_\parallel \, ,
\end{equation}
where $v^\text{p}_\parallel$ is the galaxy's peculiar velocity along the LoS (z-dimension) which is given as an output of the simulation in units of $\text{km}\,\text{s}^{-1}$. We experimented with trimming the box along the radial dimension to avoid under-dense boundary regions where galaxies have been pushed off the grid by their redefined position from equation \eqref{PlaneParaEq}. To do this we performed a $200 \, \text{Mpc}\,h^{-1}$ cut at either end so the radial box depth becomes $L_\text{z} = 600 \, \text{Mpc}\,h^{-1}$, however we found that this made no discernible difference to results for the foreground free measurements. However, we found that the depth of the box does affect the results of the foreground clean with a larger $L_\text{z}$, and therefore larger frequency range, providing a less biased foreground clean. With this in mind, we kept the angular box dimensions at $L_\text{x} = L_\text{y} = 1000 \, \text{Mpc}\,h^{-1}$ but restrict the radial depth of the box to $L_\text{z} = 762 \, \text{Mpc}\,h^{-1}$, which is equivalent to a redshift width of $\Delta z = 0.4$ at $z=0.8$ (this is approximately representative of the latest GBT measurements \citet{Switzer:2013ewa, Wolz:2015lwa}). The resolution is defined by the number of voxels where we use $N_\text{x}=N_\text{y}=N_\text{z}=225$.

Each galaxy has an associated cold gas mass $M_\text{cgm}$ and from this we can infer a \hi mass $M_\hinospace = M_\text{cgm}(1-f_\text{mol})$ where the molecular fraction is given by \citep{Blitz:2006nc}
\begin{equation}
    f_\text{mol}=\frac{R_\text{mol}}{\left(R_\text{mol}+1\right)} \, ,
\end{equation}
and we use $R_\text{mol} \equiv M_{H_2}/M_\hinospace = 0.4$ \citep{Zoldan17}. It is this \hi mass that we bin into each voxel to generate a data cube of \hi masses $M_\hinospace(\vec{r},z)$,  which should trace the underlying matter density generated by the catalogue's $N$-body simulation for the snapshot redshift $z$. These \hi masses are converted into a \hi brightness  temperature for a frequency width of $\deltadiff \nu$ subtending a solid angle $\deltadiff \Omega$ given by
\begin{equation}\label{THIequation}
    T_\hinospace(\vec{r},z) = \frac{3h_\text{P}c^2A_{12}}{32\pi m_\text{h}k_\text{B}\nu_{21}}\frac{1}{\left[(1+z)\chi(z)\right]^2}\frac{M_\hinospace(\vec{r},z)}{\deltadiff \nu \, \deltadiff \Omega} \, ,
\end{equation}
where $h_\text{P}$ is the Planck constant, $A_{12}$ the Einstein coefficient that quantifies the rate of spontaneous photon emission by the hydrogen atom, $m_\text{h}$ is the mass of the hydrogen atom, $k_\text{B}$ is Boltzmann's constant, $\nu_{21}$ the rest frequency of the 21cm emission and $\chi(z)$ is the comoving distance out to redshift $z$ (we will assume a flat universe). We refer the reader to \citet{Cunnington:2018zxg} for a more detailed discussion on equation \eqref{THIequation}.

In order to simulate this signal, we require a value for the frequency width $\deltadiff \nu$. To convert the Cartesian coordinates of the simulation box into observable frequency channels we use the comoving distance to the snapshot redshift $\chi(z)$, which we assume is the distance to the centre of the box, and the radial box length $L_\text{z}$. We then define a comoving distance to each radial bin boundary:
\begin{equation}\label{BinBoundaryEq}
    L^i_\text{z} = L^0_\text{z} + iL_\text{z}/N_\text{z}
\end{equation} 
where $i = 0,1,2, ... , N_\text{z}$ are the bin boundary indices and the distance to the minimum bin boundary is  $L^0_\text{z} = \chi(z) - L_\text{z}/2$. These can be converted into redshifts $z_i$ which are in turn converted into frequencies $\nu_i = \nu_{21} / (1+z_i)$.

In radio intensity mapping the observable signals detected by a telescope are brightness temperature fluctuations to a background mean $\overline{T}_\hinospace$, hence the observable signal is given by
\begin{equation}\label{deltaTeq}
    \deltadiff T_\hinospace(\vec{r},z) = T_\hinospace(\vec{r},z) - \overline{T}\hspace{-0.5mm}_\hinospace(z) \, .
\end{equation}
The mean \hi temperature can be related to the \hi density abundance $\Omega_\hinospace$ by \citep{Battye:2012tg}
\begin{equation}\label{TbarModelEq}
    \overline{T}\hspace{-0.5mm}_\hinospace(z) = 180\Omega_{\hinospace}(z)h\frac{(1+z)^2}{H(z)/H_0} \, {\text{mK}} \, .
\end{equation}
Constraining the \hi abundance is challenging. Whilst its value is well constrained at very low redshifts ($z \sim 0$) by targeted \hi galaxy surveys, the constraints at mid and high redshifts are few and not very competitive (see \citet{Crighton:2015pza} for a summary of available measurements in the range $0<z<5$). In principle, \hi intensity mapping with MeerKAT and the SKA should be able to provide much better constraints across a very wide range of redshifts for both the \hi abundance and the \hi bias \citep{Pourtsidou:2016dzn,Bacon:2018dui,SKAfundphys}. For example, \citet{Masui:2012zc,Switzer:2013ewa} used GBT \hi intensity mapping measurements at $z = 0.8$ (in auto and cross-correlation with WiggleZ galaxies) to measure
\begin{equation}\label{MasuiOmHI}
	\Omega_\hinospace b_\hinospace r = [4.3 \pm 1.1] \times 10^{-4} \, .
\end{equation}
Since our simulation has a finite mass resolution, it will not sufficiently sample the lowest mass halos ($\lesssim 10^{10}\,h^{-1}$M$_\odot$), which will contain \hi and therefore contribute to the intensity map. In order to ensure our simulated intensity maps have realistic amplitudes, we rescale each $T_\hinospace(\vec{r},z)$ so that it matches a model $\overline{T}_\hinospace(z)$ as per equation \eqref{TbarModelEq}. For this model we use the $\Omega_\hinospace$ constraint from equation \eqref{MasuiOmHI} setting the cross-correlation coefficient to $r=1$ and model the \hi bias with the power law \citep{Bacon:2018dui}
\begin{equation}
	b_\hinospace(z) = 0.67 + 0.18z + 0.05z^2 \, .
\end{equation}
Note that the choice of bias here is to ensure sensible values are obtained for the model $\overline{T}_\hinospace$. The degeneracy between $\Omega_\hinospace$ and $b_\hinospace$ is a further challenge for \hi intensity mapping but using RSD provides a way to break it \citep{Masui:2012zc, Pourtsidou:2016dzn}. In terms of power spectrum multipoles, this would require measuring both the monopole $P_0$ and quadrupole $P_2$, and modelling them accurately including foreground and instrumental effects -- this is the main goal of this work. 

To emulate the effects of the radio telescope beam, the observable over-temperature signal (equation \eqref{deltaTeq}) is convolved with a symmetric, two-dimensional Gaussian function with a full-width-half-maximum of $\theta_\text{FWHM}$ acting only in the directions perpendicular to the LoS. The beam size can be determined by the dimensions of the radio receiver and the redshift which is being probed. We then have
\begin{equation}\label{beamequation}
	\theta_\text{FWHM} = \frac{1.22 \, \lambda_{21} }{ D_\text{max}} (1+z) \, ,
\end{equation}
where, for single-dish intensity mapping, the maximum baseline of the radio telescope $D_\text{max}$ is simply the dish diameter. To avoid foreground cleaning problems associated with a frequency dependent beam \citep{Switzer:2013ewa, Cunnington:2019lvb}, we convolve all our maps to a constant $\theta_\text{FWHM}$ which we will explicitly state for each result.

\subsection{Simulating Foreground Contamination}\label{FGcontSec}

In order to simulate the effects of a foreground clean on our mock data we add simulated maps of known 21cm foregrounds onto our \hi cosmological signal. The first foreground we simulate is galactic synchrotron caused by electrons in the Milky Way being accelerated by the Galaxy's magnetic field. This is the most dominant foreground and can be several orders of magnitude larger than the \hi cosmological signal in the galactic plane. Extragalactic point sources (e.g. Active Galactic Nuclei) also contaminate the maps. Furthermore, free-free emission can originate both within our Galaxy and beyond causing an isotropic, extra-galactic contamination.

\begin{table}
	\centering
	\begin{tabular}{lcccr} 
		\hline
		Foreground & A & $\beta$ & $\alpha$ & $\xi$ \\
        \hline
		Galactic synchrotron & 700 & 2.4 & 2.80 & 4.0\\
		Point sources & 57 & 1.1 & 2.07 & 1.0\\
		Galactic free-free & 0.088 & 3.0 & 2.15 & 35\\
		Extra-galactic free-free & 0.014 & 1.0 & 2.10 & 35\\       
		\hline
	\end{tabular}
    \caption{Parameter values for foreground $C_\ell$ (see equation \eqref{FGpowerspec}) with amplitude $A$ given in mK$^2$. Pivot values used are $\ell_\text{ref} = 1000$ and $\nu_\text{ref} = 130 \, \text{MHz}$ as per \citet{Santos:2004ju}.}
    \label{FGparams}
\end{table}

For generating realistic foregrounds we use the Global Sky Model (GSM) \citep{deOliveiraCosta:2008pb, Zheng:2016lul} that extrapolates real data sets to provide full-sky diffuse galactic radio emission maps. Furthermore, to ensure we include contamination from all relevant foreground sources, we also use a power spectrum that is constructed to model each of the foregrounds as outlined in \citet{Santos:2004ju}. For an observation between frequency $\nu_1$ and $\nu_2$ a foreground's angular power spectrum is modelled by
\begin{equation}\label{FGpowerspec}
	C_\ell\left(\nu_1, \nu_2\right)=A\left(\frac{\ell_\text{ref}}{\ell}\right)^\beta\left(\frac{\nu_\text{ref}^2}{\nu_1\,\nu_2}\right)^\alpha \exp \left(-\frac{\log^2\left(\nu_1/\nu_2\right)}{2\,\xi^2}\right) \,,
\end{equation}
where the values for the parameters ($A$, $\beta$, $\alpha$, $\xi$) are stated in Table~\ref{FGparams} for each foreground we simulate. 

We match the frequencies of each bin to that of the \hi intensity map data by calculating a distance to each bin-boundary as defined by equation~(\eqref{BinBoundaryEq}) and then converting this into an observed redshift and frequency. We then generate a map in each frequency bin for each foreground. For the power spectrum realisation, we use 4 different random seeds (one for  each type of foreground we simulate), which we keep the same throughout each frequency bin. This ensures each foreground type has a spectral smoothness through frequency that we utilise in the foreground clean. The outputs from this and the GSM produce full-sky \texttt{HEALPix}\footnote{\href{http://healpix.sourceforge.net}{https://healpix.sourceforge.io/}} \citep{2005ApJ...622..759G,Zonca2019} maps but we convert them into flat-sky Cartesian maps in order to add to our \hi data. To do this we define an angular coordinate for each pixel on the flat-sky map, which we match to a pixel in the \texttt{HEALPix} map with the closest angular coordinate. While this approach is an approximation and may affect some angular coherence in the foreground maps, it will have no impact on the foreground as a contaminant to our data. We chose to match the centre of the flat-sky maps to the centre ($\text{right-ascension} = \text{declination} = 0\,\text{deg}$) of the \texttt{HEALPix} map. This means our foreground data is coming directly from the centre of the galactic plane. In reality, it is likely that this region will be avoided, since this is where foregrounds are expected to be strongest. However, in order to maximize foreground amplitudes and ensure as robust a test as possible, it is from these regions we chose to cut a patch, equal to the size of our \hi data coverage.

\begin{figure}
	\centering
  	\includegraphics[width=\columnwidth]{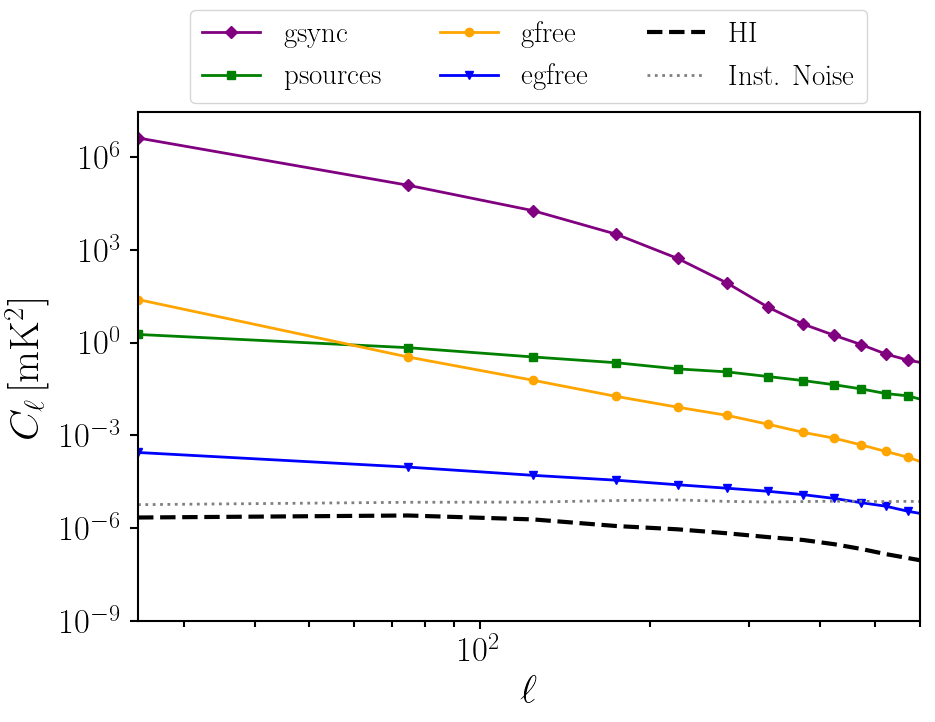}
    \caption{Measured angular power spectra for each component of the observed signal at $\nu = 780\,\text{MHz}$ ($z\sim 0.8$) with a dish diameter of $100\,\text{m}$. Foregrounds are shown as solid coloured lines. The \hi cosmological signal produced using the \multidark\ simulation for a $z=0.82$ snapshot is the black dashed line. The grey dotted line shows the contributions from the instrumental noise (see Section~\ref{NoiseSec}).}
\label{FGPk}
\end{figure}

The different foreground types are added and the frequency slices are stacked to form the foreground data cube with the same [$N_\text{x}, N_\text{y}, N_\text{z}$] structure, which we then add onto our $\delta T_\hinospace$ data cube. These foregrounds dominate over the cosmological signal by many orders of magnitude as we show in Figure~\ref{FGPk}. These are the measured power spectra for each foreground and the simulated cosmological signal (black dashed line) with all signals at an observed frequency of $\nu = 780$ MHz ($z\sim 0.8$), with a dish diameter of $D_\text{max}=100\,\text{m}$.

In order to recover the useful cosmological signal we therefore require a foreground removal process. For this we use Fast Independent Component Analysis (\fastica) \citep{Hyvrinen1999FastAR}. This is a blind foreground removal method where we assume that a raw observed signal, such as that outlined in equation \eqref{deltaTeq}, can be generalized into a linear equation where the elements making up the signal are statistically independent. Therefore for each LoS, sorted into $N_z$ redshift bins and assuming $m$ independent components are present, \fastica\ assumes the observed signal can be written as
\begin{equation}\label{ICAequation2}
	\textbf{x} = \textbf{A}\textbf{s} + \varepsilon = \sum_{i=1} ^{N_\text{IC}=m} \textbf{a}_i s_i + \varepsilon \, ,
\end{equation}
where $\textbf{s}$ are the $m$ independent components, $\textbf{A}$ is the mixing matrix determining the amplitudes of the independent components and $\varepsilon$  is the residual (containing \hi signal and noise). The number of independent components $m$ is an input and we find that too low a value causes higher foreground residuals but a very high value starts to damage the signal at low-$k$ \citep{Alonso:2014dhk}. For this work we use $m=4$ for all our \fastica\ foreground cleans, finding this to be sufficient and a commonly used value in previous work \citep{Chapman:2012yj, Wolz:2013wna}. The residual $\varepsilon$ should therefore contain the cosmological information we require and it is this residual that we will refer to as our cleaned data. We will make comparisons between this cleaned case and the idealised case where maps are completely free of foregrounds. \fastica\ is capable of removing the simulated foregrounds across a wide range of scales but, as one might predict, at low-$k$ there is some discrepancy -- that is because the foregrounds are smooth in frequency and thus largely exist in the small $k_\parallel$ modes. This makes cosmological signal in this region of $k$-space hard to disentangle from the foregrounds and thus some signal is lost due to the foreground cleaning. For a more detailed discussion on foregrounds and their subtraction using \fastica\ for \hi intensity mapping we refer the reader to \citet{Wolz:2013wna,Alonso:2014dhk,Cunnington:2019lvb}.

\subsection{Instrumental Noise}\label{NoiseSec}

Along with foregrounds, we also need to take into account the instrumental (thermal) noise from the radio telescope. In order to add this instrumental noise to our data cubes we simulate uncorrelated Gaussian fluctuations. These are added after the beam is applied, which means that the final noise fluctuations remain uncorrelated. While this would not be true in a realistic situation, we opt for this simpler approach for our purposes since we also assume a perfectly Gaussian beam of known size. Hence, we add onto the observable maps a Gaussian random field with a spread given by
\begin{equation}\label{NoiseEq}
    \sigma_\text{noise} = T_\text{sys}\, \sqrt{\frac{4\pi\, f_\text{sky}}{\Omega_\text{beam}\,N_\text{dish}\,t_\text{obs}\,\deltadiff\nu}} \, .
\end{equation}
Here $T_\text{sys}$ is the total system temperature, which is the sum of the sky and receiver noise \citep{Santos:2015bsa}; $\Omega_\text{beam} \simeq 1.133\theta_\text{FWHM}^2$ is the solid angle for the intensity mapping beam; $f_\text{sky}$ is the fraction of sky covered by the survey, which for a box with perpendicular size of $1000\,\times\,1000\,\text{Mpc}^2\,h^{-2}$ at $z=0.82$ is approximately $29\,\times\,29\,\text{deg}^2$ and therefore $f_\text{sky}\sim 0.02$. In order to achieve realistic levels of noise that slightly dominate over the \hi signal, as is expected in near future intensity mapping experiments, we use one dish ($N_\text{dish}=1$), set $T_\text{sys}=10\,\text{K}$ and assign a total observation time $t_\text{obs}=200$ hours. The instrumental noise power spectrum using these specifications is shown in Figure~\ref{FGPk} as the grey dotted line.

A technique used in the GBT intensity mapping observations \citep{Switzer:2013ewa,Masui:2012zc,Wolz:2015lwa} is to create data sub-sets by observing the same patch of sky (field) at different times, such that the instrumental noise is independent in each sub-set map. These sub-sets can then be cross-correlated to suppress the noise and give the \hi auto-correlation signal (while the sub-sets auto-correlation can be used as a proxy for the noise itself). We emulate this approach by simulating two independent noise maps using equation \eqref{NoiseEq} and create two observable \hi intensity map sub-sets with the same underlying cosmological signal but independent instrumental noise. These are then cross-correlated to produce our \hinospace-auto-correlation result.

As discussed in Section~\ref{ModellingSec}, there is also the contribution of shot noise. Fortunately for \hi intensity mapping this value is expected to be fairly subdominant \citep{Spinelli:2019smg}. Using the above formalism we will aim to replicate an observational experiment and include shot noise in our modelling in line with equations \eqref{ObsMultipole0}, \eqref{ObsMultipole2} and \eqref{ObsMultipole4}, using an external result from a simulation. For this we use results from \citet{Villaescusa-Navarro:2018vsg} and adopt their values of $P_\text{SN}=\overline{T}^2_\hinospace 124\,\text{mK}^2\,\text{Mpc}^3\,h^{-3}$ for our $z\sim 0.8$ results and $P_\text{SN}=\overline{T}^2_\hinospace 65\,\text{mK}^2\,\text{Mpc}^3\,h^{-3}$ for $z\sim 2$.

There are other forms of noise associated with \hi intensity mapping such as correlated 1/$f$ noise \citep{Bigot-Sazy:2015jaa} and Radio Frequency Interference (RFI) noise of which Global Navigation Satellite Services (GNSS) have been identified as a potentially big problem \citep{Harper:2018ncl}. In this work, we assume these issues are controllable or mitigated.

\section{Results}\label{ResultsSec}

Here we present the results from our simulation measurements, which demonstrate the foreground and instrumental effects on the \hi intensity mapping power spectrum multipoles. For all plots demonstrating the effect of foreground contamination we use a consistent convention where black dashed lines and black cross data points represent the \textit{foreground free} case. Whereas red dotted lines and red circle data points represent the \textit{foreground contaminated} case. Foreground contaminated case refers to 21cm foregrounds being added to the simulation and then cleaned using a \fastica\ reconstruction as outlined in Section~\ref{FGcontSec}. Where applicable we include error bars, which for the power spectrum multipoles are given by \citep{Feldman:1993ky,Seo_2010,Battye:2012tg, Grieb:2015bia, Blake:2019ddd}: 
\begin{equation}\label{powerspecerror}
	\sigma_{P_\ell}(k) = \frac{(2\ell+1)}{\sqrt{N_\text{modes}}}\,\sqrt{\int_{0}^{1} \diff\mu\,(P(k,\mu)+P_\text{N})^2\mathcal{L}_\ell^2}\, ,
\end{equation}
where $N_\text{modes}$ is the number of unique modes in the bin and the noise power spectrum can be written as $P_\text{N} = \sigma_\text{noise}^2V_\text{vox}$. We have assumed a Gaussian diagonal covariance and we refer the reader to B19 for a more comprehensive discussion of errors in this context. We also tested using a jackknifing process for the error bars and found good agreement between the two approaches. We include both options for generating error bars in the accompanying \texttt{python} toolkit.

Unless otherwise stated we use the \multidark\ simulation for our \hi data, which are used to generate intensity maps using the methods outlined in Section~\ref{CosmoSigSec}. We mostly chose to run our results at a redshift of $z=0.82$, which is representative of the redshifts that current and forthcoming single-dish intensity mapping experiments are targeting, for example the GBT intensity mapping survey \citep{Switzer:2013ewa}. We assume a dish size of $100\,\text{m}$, which is chosen to give a relatively small beam size $R_\text{beam}=3.9\,\text{Mpc}\,h^{-1}$ (from equation \eqref{beamequation}). This is to allow the effects from the foregrounds on the simulated data to be more evident. We initially run our results without instrumental noise, again as this demonstrates the effect of foregrounds with more clarity. We then investigate the effects of adding in realistic noise as introduced in Section~\ref{NoiseSec} and show results with a larger beam.

\subsection{Demonstrating the Observational Effects}

\begin{figure*}
	\centering
  	\includegraphics[width=2\columnwidth]{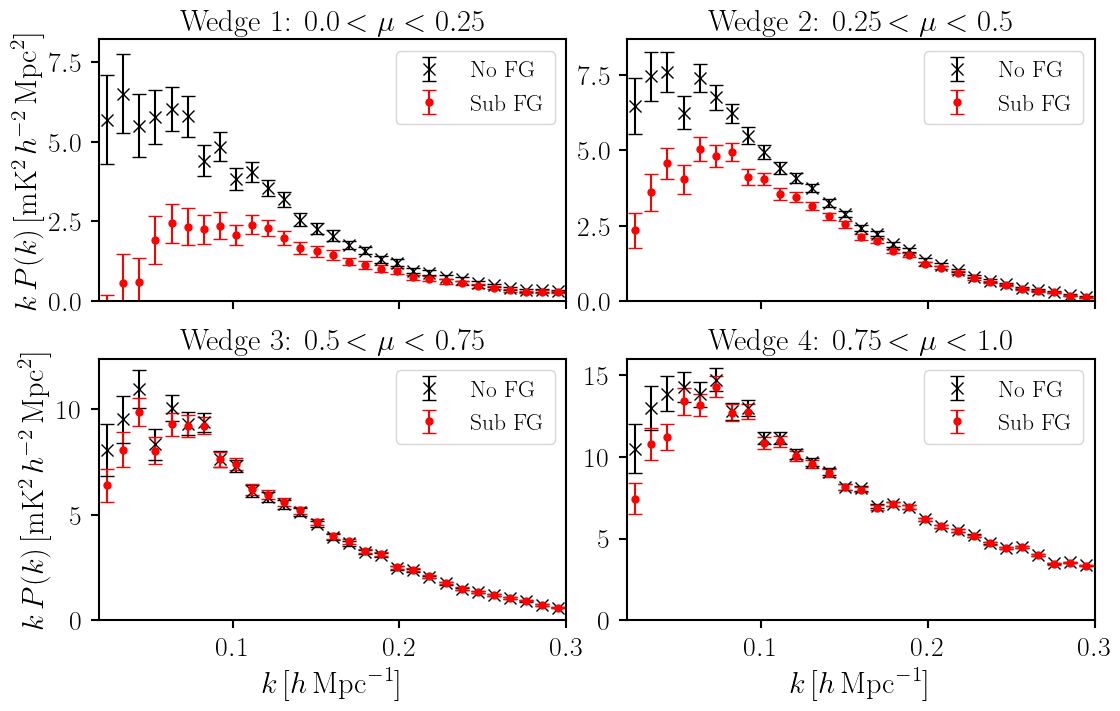}
    \caption{Power spectrum $P^{\rm wedge}(k)$. The different wedges are calculated following equation \eqref{eq:Pwedge}. For each wedge we show the differences between the case without foregrounds (black cross points) and where foregrounds are added then removed with \fastica\ (red circle points). Produced with the \multidark\ simulation at $z=0.82$ with a $R_\text{beam} = 6\,\text{Mpc}\,h^{-1}$ beam size.}
\label{muwedgeFGplot}
\end{figure*}

The clustering wedges, as introduced in Section~\ref{WedgesSec} and defined by equation~\eqref{eq:Pwedge}, have been used as a means for avoiding foreground contaminated regions
in past work \citep{Raut:2017zhh}. Building upon this idea, it is interesting to look at results from our simulations in different clustering wedges and focus mainly on how foregrounds are having an impact.

Figure~\ref{muwedgeFGplot} shows the effect a foreground clean has on different power spectrum wedges. As shown, we choose four different wedge bins spanning the full range of angles to the LoS. This demonstrates why the idea of a $\mu$ cutoff can be useful in the context of foreground contamination. Since $\mu$ is the directional cosine of the modes and therefore $k_\parallel = k\mu$, a wedge with only low-$\mu$ included (as in the \textit{top-left} plot of Figure~\ref{muwedgeFGplot}) means only small $k_\parallel$ modes are included and these are the ones most affected by the foregrounds. The extreme case of the lowest-$\mu$ wedge clearly demonstrates the effect foregrounds have on the power spectrum, with power being drastically damped for low-$k$ modes. The results in Figure~\ref{muwedgeFGplot} also show that the impact of foregrounds becomes less severe as we move towards higher $\mu$ regions, as expected.

In the last wedge (\textit{bottom-right}) where the largest $\mu$-values are displayed we can see little effect from foregrounds. This wedge also demonstrates the effect of the telescope beam, which is also heavily dependent on the wedge used. To emphasise the effects from the beam, we use a slightly larger beam than the default $100\,\text{m}$ dish at $z=0.8$  and instead use a beam with $R_\text{beam} = 6\,\text{Mpc}\,h^{-1}$ for these clustering wedges. In comparison with the other wedges, the power in the high-$\mu$ wedge is larger at high-$k$. That is because the high-$k_\perp$ modes damped by the beam have been excluded in this wedge. Referring back to Figure~\ref{muMap} we can see that $\mu>0.75$ represents mostly small-$k_\perp$ modes, which are unaffected by the beam (since it damps smaller perpendicular scales, i.e. large-$k_\perp$). However, for the other wedges, the modes can be composed of higher $k_\perp$ contributions that the beam effectively damps.

\begin{figure*}
	\centering
  	\includegraphics[width=2.1\columnwidth]{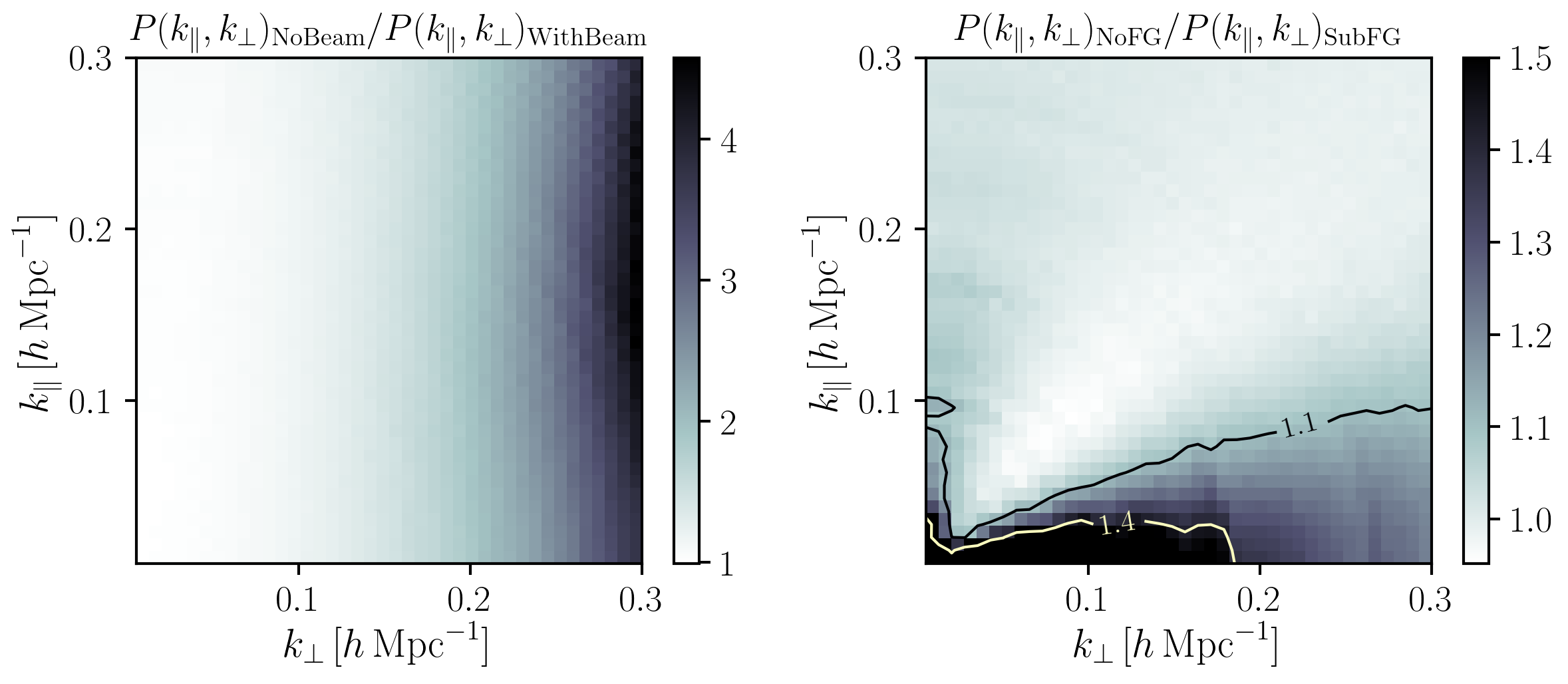}
    \caption{The impact of observational effects on the power spectrum decomposed into parallel ($k_\parallel$) and perpendicular ($k_\perp$) modes. \textit{Left}-panel demonstrates the impact of a $R_\text{beam}=3.9\,\text{Mpc}\,h^{-1}$ telescope beam by showing the ratio of the power spectrum for foreground-free \multidark\ intensity maps both with and without smoothing to emulate the beam. \textit{Right}-panel shows the difference between foreground free and foreground cleaned intensity maps. Both results are at $z=0.82$ and in the foreground comparison we smoothed the maps to emulate the beam for both $P_\text{NoFG}$ and $P_\text{SubFG}$ cases for consistency and use $N_\text{IC}=4$ for the \fastica\ foreground removal. For the foreground case (\textit{right}-panel), to avoid saturation i.e. to avoid the smallest $k_\parallel$ modes showing a difference $\gg 1$, we have limited the maximum value of $P_\text{NoFG}/P_\text{SubFG}$ to 1.5.}
\label{kperpkpara}
\end{figure*}

The clustering wedges nicely demonstrate the anisotropic nature of the foreground removal and beam effects, however they do not completely disentangle the contributions from parallel ($k_\parallel$) and perpendicular ($k_\perp$) modes. To make this clearer, in Figure~\ref{kperpkpara} we split these contributions explicitly, i.e. we study $P(k_\perp,k_\parallel)$. The \textit{left} panel of Figure~\ref{kperpkpara} shows the difference between an intensity map with and without a beam. We can see that the biggest differences (darker-regions) occur for the highest-$k_\perp$ modes and this demonstrates, as expected, how the damping from the telescope beam is $k_\perp$-dependent.

The \textit{right} side panel shows a comparison between foreground free intensity maps and ones that have been contaminated with foregrounds and then cleaned with \fastica. Again, as expected, this shows that the effects from foregrounds are largely a function of $k_\parallel$ with the smallest parallel modes being most affected. However, the plot reveals that there is also a $k_\perp$ dependence and in fact, comparison with Figure~\ref{muMap} reveals that the effect of foreground removal has some strong $\mu$-dependence. It is visible by eye how the most affected regions from foregrounds correspond to the area in $\mu$-space where $\mu \lesssim 0.2$. This interesting result suggests that if, for example, we aim to construct an estimator that avoids the most contaminated modes, then just parameterizing this using $k_\parallel$ only, may not be the most optimal approach (at least in the context of this simulation and with this particular foreground cleaning method). 
It is plausible that the angular structure that we see in Figure \ref{kperpkpara} is a result of the the angular structure in the foregrounds. We find slightly less angular structure in these power spectrum residuals if we remove the galactic synchrotron map from the simulations. This map is the only one extrapolated from real data and has higher intensity closer to the galactic plane. The other foregrounds are simulated from Gaussian realizations (as discussed in Section \ref{FGcontSec}) and therefore have less angular structure. Further investigation of these results is beyond the scope of this paper and is left for future work.

\begin{figure}
	\centering
  	\includegraphics[width=\columnwidth]{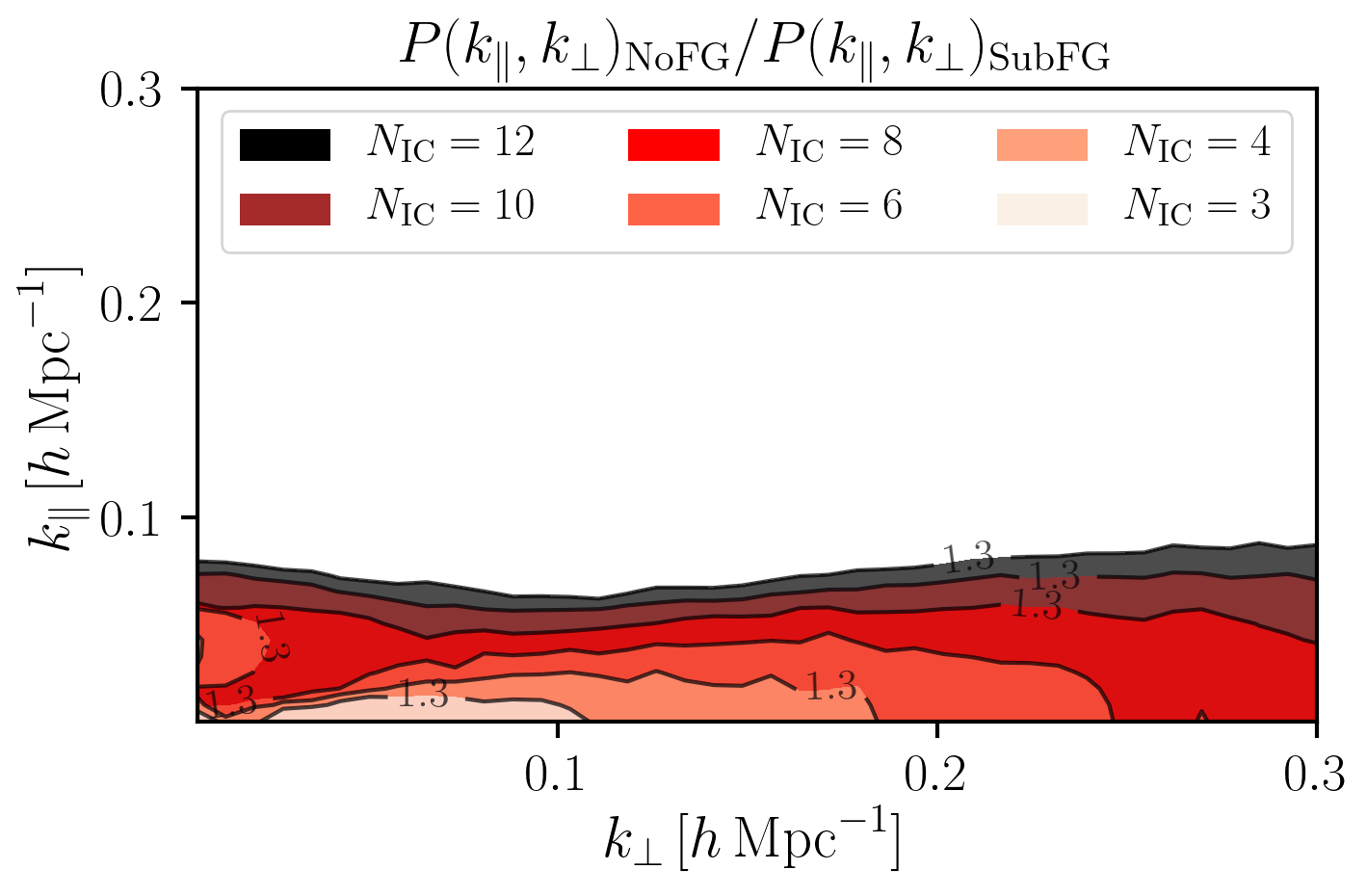}
    \caption{Same as \textit{right} panel of Figure~\ref{kperpkpara} but here we only show the $P_\text{NoFG}/P_\text{SubFG}=1.3$ contours for different numbers of independent components ($N_\text{IC}$) chosen for the \fastica\ foreground removal. The coloured regions mark where the ratio of the foreground free and subtracted foregrounds is greater than 1.3, i.e. the most foreground affected regions.}
\label{NICcontours}
\end{figure}

We also demonstrate in Figure~\ref{NICcontours}, that when the number of independent components ($N_\text{IC}$) chosen in the \fastica\ foreground reconstruction is increased, then the loss of modes from foregrounds does become more consistent and better approximated by a low-$k_\parallel$ cut. It is plausible that the number of independent components we consistently use throughout this work ($N_\text{IC}=4$) will not be sufficient for real data, which will likely require higher $N_\text{IC}$, as explored in \citet{Wolz:2015lwa} using the GBT observations. This will especially be true when dealing with more stubborn foregrounds from polarization leakage, which we have not simulated in this work. This will probably need a much more aggressive foreground removal approach and setting $N_\text{IC} \gtrsim 10$ might be required, at the cost of damping more \hi cosmological signal.
 
In Figure~\ref{NICcontours} we demonstrate the effect of increasing the number of independent components. This shows that using a higher number of independent components $N_\text{IC} \gtrsim 10$, which is more likely when dealing with real data, the foreground contaminated modes become more easily defined as a simple low-$k_\parallel$ region. This means constructing an estimator that utilizes a foreground avoidance method might be an easier task and only requires some definition of a $k^\text{FG}_\parallel$ cutoff below which data are excluded. Confirming this claim would be an interesting investigation but would need the inclusion of simulated polarization leakage. We leave this investigation for a follow-up study and stick to using the default $N_\text{IC} = 4$ for the rest of this work. 

\subsection{Null-RSD Test}\label{NoRSDSec}

\begin{figure*}
	\centering
  	\includegraphics[width=2.1\columnwidth]{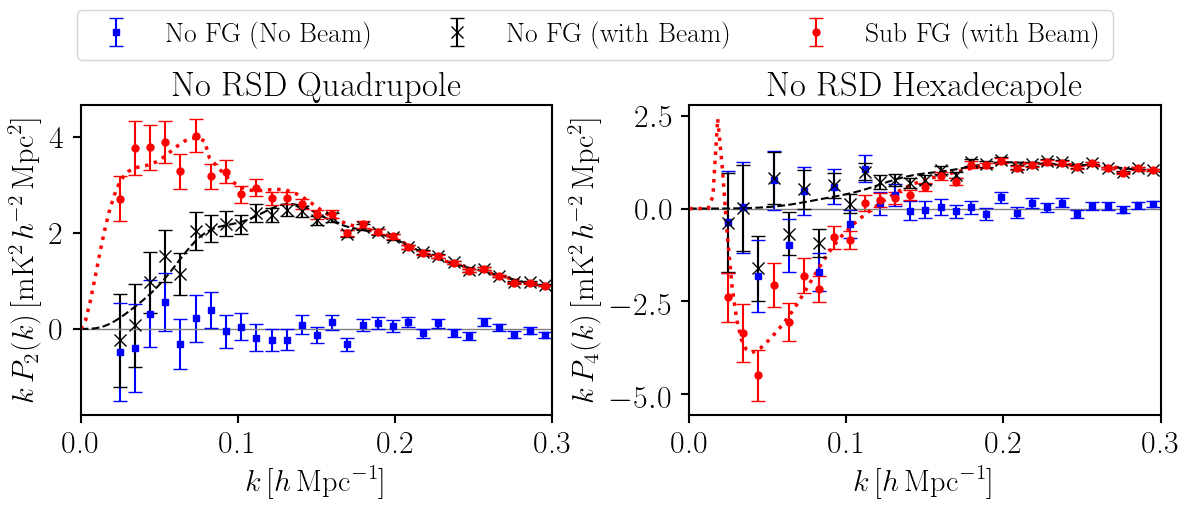}
    \caption{Quadrupole ($P_2$) and hexadecapole ($P_4$) for auto-correlations of \hi intensity maps, produced using \multidark\ simulations with no RSD at $z=0.82$ with a $R_\text{beam} = 10\,\text{Mpc}\,h^{-1}$ beam size. The results should be $P_2 = P_4 = 0$ due to the exclusion of RSD from the simulation that the foreground and beam free results show (blue square points). Introducing a beam (black cross points) and then also introducing foreground contamination (red circle points) creates a non-zero signal in both multipoles. However, this can be modelled as demonstrated by the agreement with the predictions shown by the dashed and dotted lines, which represent the $\hat{P}_\ell$ model (equation \eqref{ToyFGremovalEq}) but without Kaiser and FoG factors.}
\label{NoRSDMultipole}
\end{figure*}

Since a \hi intensity mapping quadrupole (and hexadecapole) detection would correspond to a detection of RSD in the radio wavelength, it is very important to understand and model the anisotropic effects of foreground removal and the instrumental beam. As our toy model showed, what might appear as a detection could just be systematics interacting with the Legendre polynomials to create a false enhanced signal. To investigate this in more detail we have used the \multidark\ simulation as done in the previous examples in this section, but instead removed RSD; this is simply done by not displacing the galaxy positions along the LoS, i.e. by not including the peculiar velocities contribution in equation \eqref{PlaneParaEq}. This should result in a null quadrupole and hexadecapole. However, we find that in the presence of the telescope beam and foreground contamination, a false signal appears.

Figure~\ref{NoRSDMultipole} shows both the quadrupole (\textit{left} plot) and the hexadecapole (\textit{right} plot) for this null RSD test. In each case we show the measurement with no beam and no foregrounds (blue square data points) and as expected we get a null signal. However, when we smooth the intensity maps to emulate the effect of the telescope beam we begin to see some non-zero signal mostly in the higher-$k$ range (black cross data points). To emphasise the effects from the beam, we use $R_\text{beam} = 10\,\text{Mpc}\,h^{-1}$ for the beam size. We then also introduce effects from the foregrounds by adding on simulated foregrounds maps and cleaning them with \fastica\ (red circle data points). As shown, introducing foregrounds creates a non-zero signal this time mostly in the lower-$k$ range.

We also show the $\hat{P}_2$ and $\hat{P}_4$ (equation \eqref{ToyFGremovalEq}) model predictions for these cases as the black dashed line (no foregrounds) and red dotted line (with foregrounds), which in this case has been calculated in the same way as outlined in Section~\ref{ModellingSec} but with the Kaiser and FoG factors excluded. Interestingly, for the quadrupole in the foreground subtracted case, we achieve better agreement with the data for low-$k$ if we use a constant $\mu_\text{FG}$ cut rather than the $\mu_\text{FG} = k_\parallel^\text{FG}/k$ varying parameter from equation \eqref{muFGeq}. The results in Figure~\ref{NoRSDMultipole} therefore use a $\mu_\text{FG}=0.22$ cut for $\hat{P}_2$ with $k<0.08\,h\,\text{Mpc}^{-1}$ but stick with using the equation \eqref{muFGeq} cut with a best fit value of $k_\parallel^\text{FG}=0.015\,h\,\text{Mpc}^{-1}$ for the rest of the $k$ values, and for all $\hat{P}_4$. This result is further confirmation that the effects from foreground removal are not purely $k_\parallel$ dependent in the context of this non-aggressive ($N_\text{IC}=4$) \fastica\ clean, as we also saw in the right-panel of Figure~\ref{kperpkpara}.

To test the agreement with the model, we calculate the reduced $\chi^2$ statistic, which is $\chi^2/\text{dof}=1.29$ when we just consider the effects from the beam. This rises to $\chi^2/\text{dof}=1.76$ when the beam and foreground effects are included. This latter $\chi^2$ result suggests that the data is unlikely to be drawn from the model, however, in this test we have not included instrumental noise that would increase the errors and thus improve the reduced $\chi^2$ statistic. Furthermore, under the effects of both the beam and foregrounds the quadrupole data only has an average $\sim5\%$ deviation from the model, which reveals a decent agreement. Results are worse for the hexadecapole where signal-to-noise is expected  to be lower, but the general trends predicted by the model are still followed.

We emphasize that the data has also been integrated across the full range of $\mu$ and thus our model assumes the foreground affected $P(k,\mu)$ is zero in the $\mu < \mu_\text{FG}$ regions and unaffected otherwise. This model is an idealised description of the foreground clean and is likely the reason for the slightly high $\chi^2$ statistic. However, if aiming to optimise agreement to a model, e.g. for parameter estimation (as we begin to investigate in Section \ref{MCMCSec} and aim to follow-up in further work), then also limiting the range of $\mu$ when evaluating the multipoles for the data, would achieve a more consistent comparison with the model and safeguard against $P(k,\mu)$ not being exactly zero for all $\mu<\mu_\text{FG}$. 

The results from Figure \ref{NoRSDMultipole} largely confirm our understanding for the source of these signals, which is due to the anisotropic nature of the  beam and foreground removal effects weighted by the Legendre polynomials in the multipole expansion. The combined effect from the telescope beam and foreground contamination results in a non-zero signal. This means that when working with real \hi intensity mapping data, we need to be very confident in our modelling and control of the foreground removal and beam effects on the different multipoles, especially when we aim to measure RSD.

\subsection{Foreground Effects on the Multipoles}

\begin{figure*}
	\centering
  	\includegraphics[width=2\columnwidth]{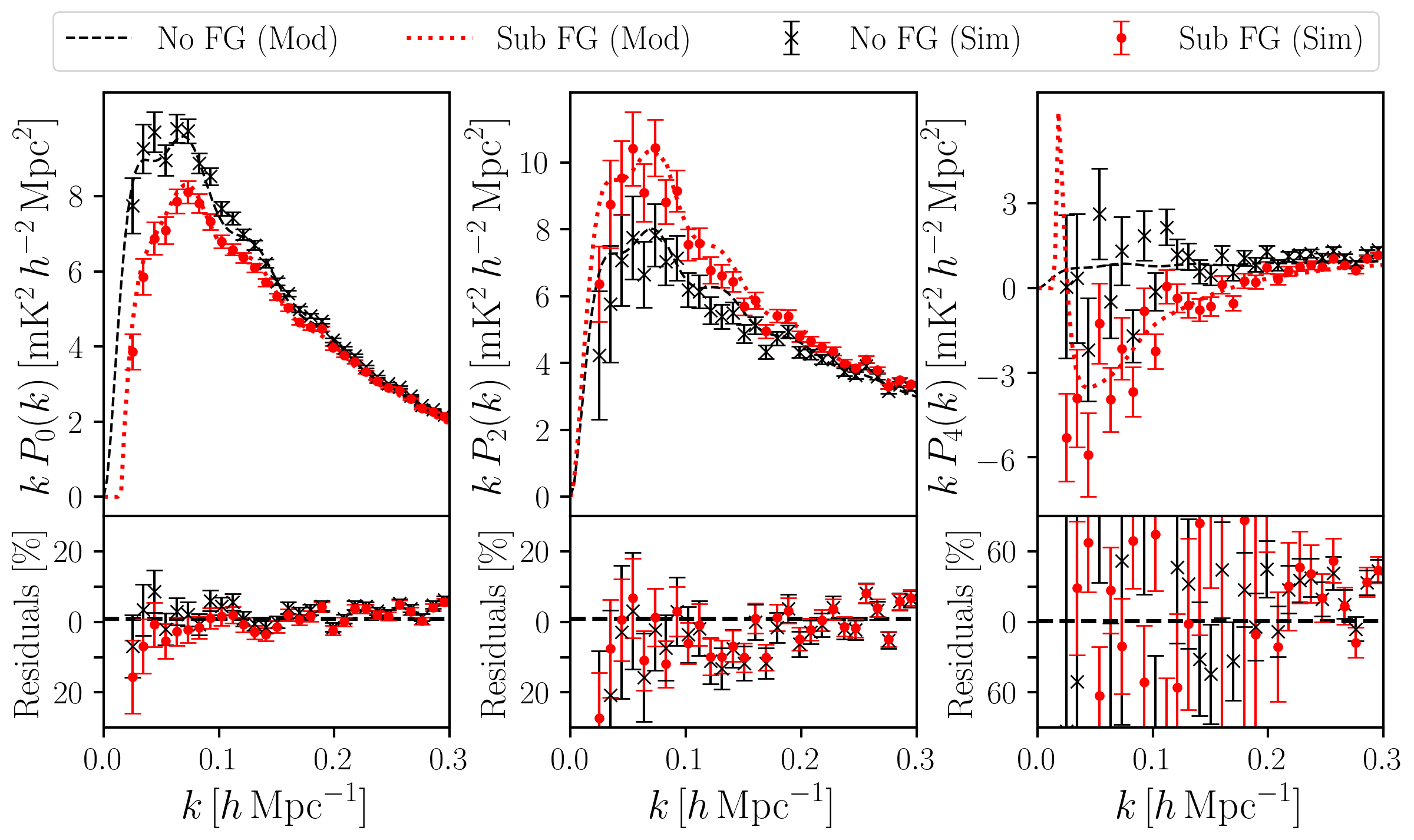}
	\caption{Power spectrum multipoles using the simulated \multidark\ intensity maps at redshift of $z=0.82$ with a $R_\text{beam} = 3.9\,\text{Mpc}\,h^{-1}$ beam size but no instrumental noise. Bottom panels show the percentage residual difference between the simulated data and the model $\hat{P}_\ell$ (predicted by equation \eqref{ToyFGremovalEq}) with $\mu_\text{FG}=0$ for the No FG cases (black dashed) and $\mu_\text{FG}=k^\text{FG}_\parallel/k$ (equation \eqref{muFGeq}) for the Sub FG cases (red dotted).}
	\label{MDSAGEMultipolesz0p8}
\end{figure*}

To demonstrate and explain the impact that foregrounds have on the amplitude and shape of each of the multipoles, it is useful to perform a first test without any instrumental noise for clarity. Furthermore, with our choice of large $100\,\text{m}$ dish, the beam should be sufficiently small at $z=0.82$ that the effects of foregrounds should be clear as demonstrated in Section~\ref{BigBeamModelSec}. Figure~\ref{MDSAGEMultipolesz0p8} represents this first measurement of the three multipoles ($P_0$, $P_2$ and $P_4$) for the \multidark -\textsc{Sage} simulated maps with RSD effects applied. For each power spectra we show the foreground free maps (black cross data points) along with maps with foregrounds added and then cleaned with \fastica\ (red circle data points). We show our fiducial $\hat{P}_\ell$ as introduced in Section~\ref{ModellingSec}, which attempts to model the effects of foregrounds by limiting the $\mu$ range integrated over by assuming all modes with $\mu<\mu_\text{FG}$ are lost (red dotted lines). The foreground free case (black dashed lines) assumes $\mu_\text{FG}=0$. 

For these results we used a bias of $b_\hinospace = 1.15$ and  $\sigma_\text{v}=250\,h^{-1}\,\text{km}\,\text{s}^{-1}$ to achieve the best agreement with the fiducial model, which is approximately consistent with predictions of the bias at this redshift \citep{Villaescusa-Navarro:2018vsg,Spinelli:2019smg}. In general the agreement with this fiducial model is good, especially in the monopole as shown by the residuals in the bottom panel, which average $\sim3\%$ deviation from the model. In the monopole we see a damping from foregrounds that gets more severe for lower-$k$ as predicted. We see higher errors in the quadrupole as would be expected since signal to noise is expected to be worse the higher the multipole. We perhaps see some slight discrepancies in the quadrupole as well where the model appears to overestimate the power in the mid-range of the $k$-values. The overall effect from foregrounds is still consistent with what is predicted in that we see an enhanced quadrupole when foregrounds are subtracted. Within the high scatter, the hexadecapole results also appear to agree with the trend predicted by the model, albeit less conclusively due to the lower signal-to-noise ratio as demonstrated by the large error bars and residual percentages in the bottom panel.
 
The inclusion of the fiducial model is mainly to demonstrate that the overall anisotropic observational effects can be approximated. A more robust fit to the data will have to include a full fitting analysis and exploration of different (e.g. perturbative) models. But it is encouraging that a decent agreement can still be obtained with this fiducial model. The difference in model agreement between the foreground and foreground free cases appears minimal, which is encouraging. Similarly to the results in Figure \ref{NoRSDMultipole}, for the quadrupole in the foreground subtracted case, we achieve better agreement with the data for low-$k$ if we use a constant $\mu_\text{FG}=0.16$ cut for $\hat{P}_2$ with $k<0.08\,h\,\text{Mpc}^{-1}$ but stick with using the equation \eqref{muFGeq} cut with a best fit value of $k_\parallel^\text{FG}=0.015\,h\,\text{Mpc}^{-1}$ for the rest of the $k$ values, and for all $\hat{P}_0$ and $\hat{P}_4$.

As discussed in the previous Section \ref{NoRSDSec},
for the foreground affected case, the model is assuming $P(k,\mu)$ is exactly zero for all $\mu<\mu_\text{FG}$, which is probably an unrealistic assumption. If the priority was to optimise agreement to a model, then the multipoles from the data could be measured in the same restricted $\mu$ range to be consistent with the model. This would safeguard against the imperfect assumption that all modes are lost with $\mu<\mu_\text{FG}$ because we would be excluding them. However, this then becomes less demonstrative and we could for example set quite a high cut in $\mu_\text{FG}$, achieve good model agreement, but not learn as much on the extent of the impact from the foreground clean. In this work our priority is to demonstrate the effect of foregrounds and we therefore chose to evaluate the multipoles for the simulated data by integrating over the full range of $\mu$ and then attempt to match a model to these results.

\begin{figure*}
	\centering
  	\includegraphics[width=2\columnwidth]{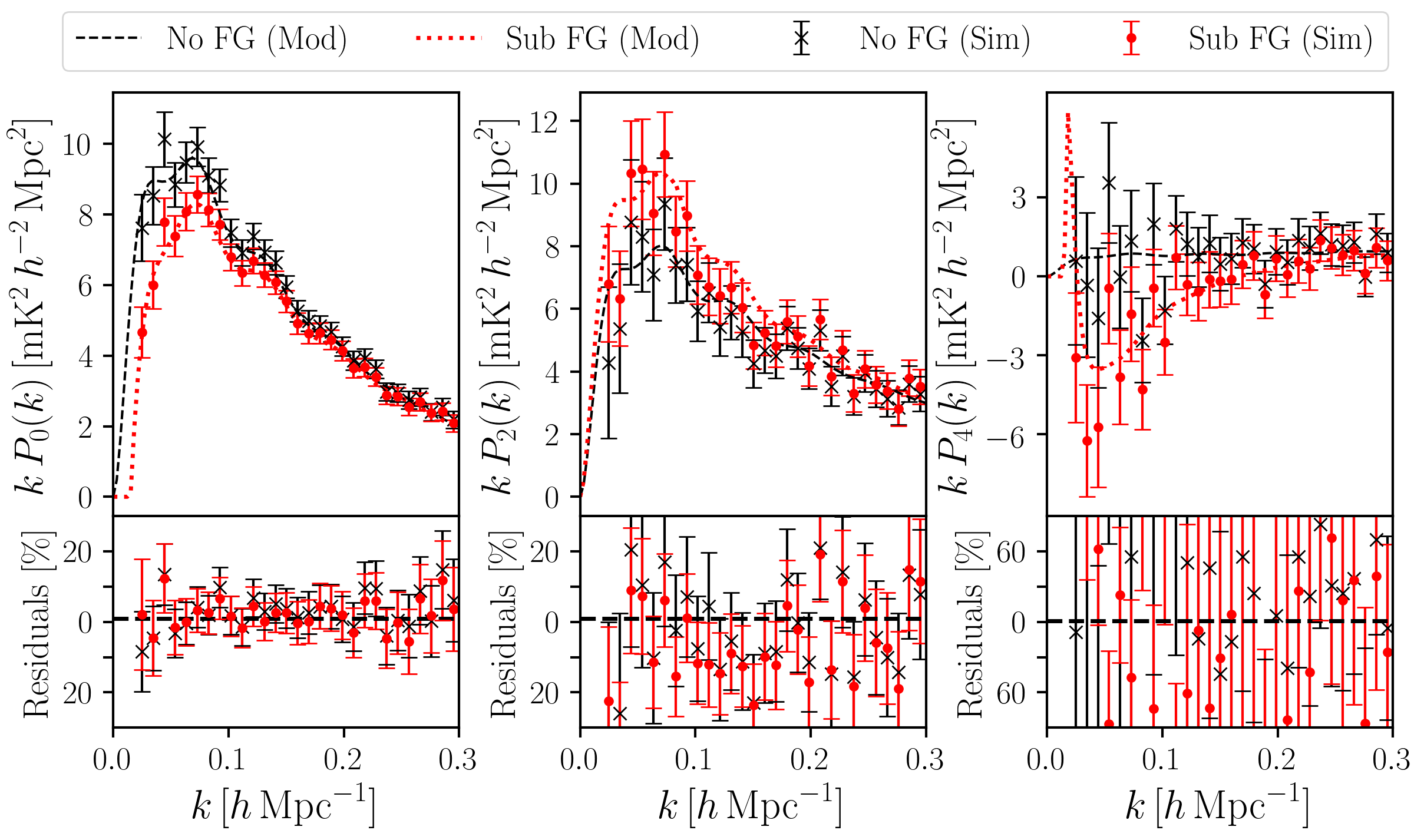}
	\caption{Same plot as Figure~\ref{MDSAGEMultipolesz0p8} but with dominant instrumental noise included in the simulations. An increase in error bar size and residual percentage errors (bottom panel) is evident from the noise introduction.}
	\label{MDSAGEMultipolesz0p8wNoise}
\end{figure*}

Figure~\ref{MDSAGEMultipolesz0p8wNoise} shows the same test but in the presence of instrumental noise. As demonstrated by Figure \ref{FGPk}, we include instrumental noise that dominates over the \hi signal across most scales. We test the method outlined in Section~\ref{NoiseSec}, which cross-correlates different sub-sets of \hi intensity maps, each with independent instrumental noise. As shown in Figure~\ref{MDSAGEMultipolesz0p8wNoise}, despite these high levels of instrumental noise, each multipole is still measured to be in general agreement with the model predictions. There are some obvious differences with respect to the results of Figure~\ref{MDSAGEMultipolesz0p8}. Firstly, the errors are noticeably larger and the residual percentage differences between model and data have increased. This is unsurprising in the presence of the instrumental noise, which dominates over the \hi signal, especially at high-$k$. In this more realistic test we achieve a reduced $\chi^2$ statistic of $\chi^2/\text{dof}\sim1$ for all multipoles in both foreground and foreground free cases. 

\subsection{Higher Redshift \& Larger Beam}\label{HighzResults}

\begin{figure*}
	\centering
  	\includegraphics[width=2\columnwidth]{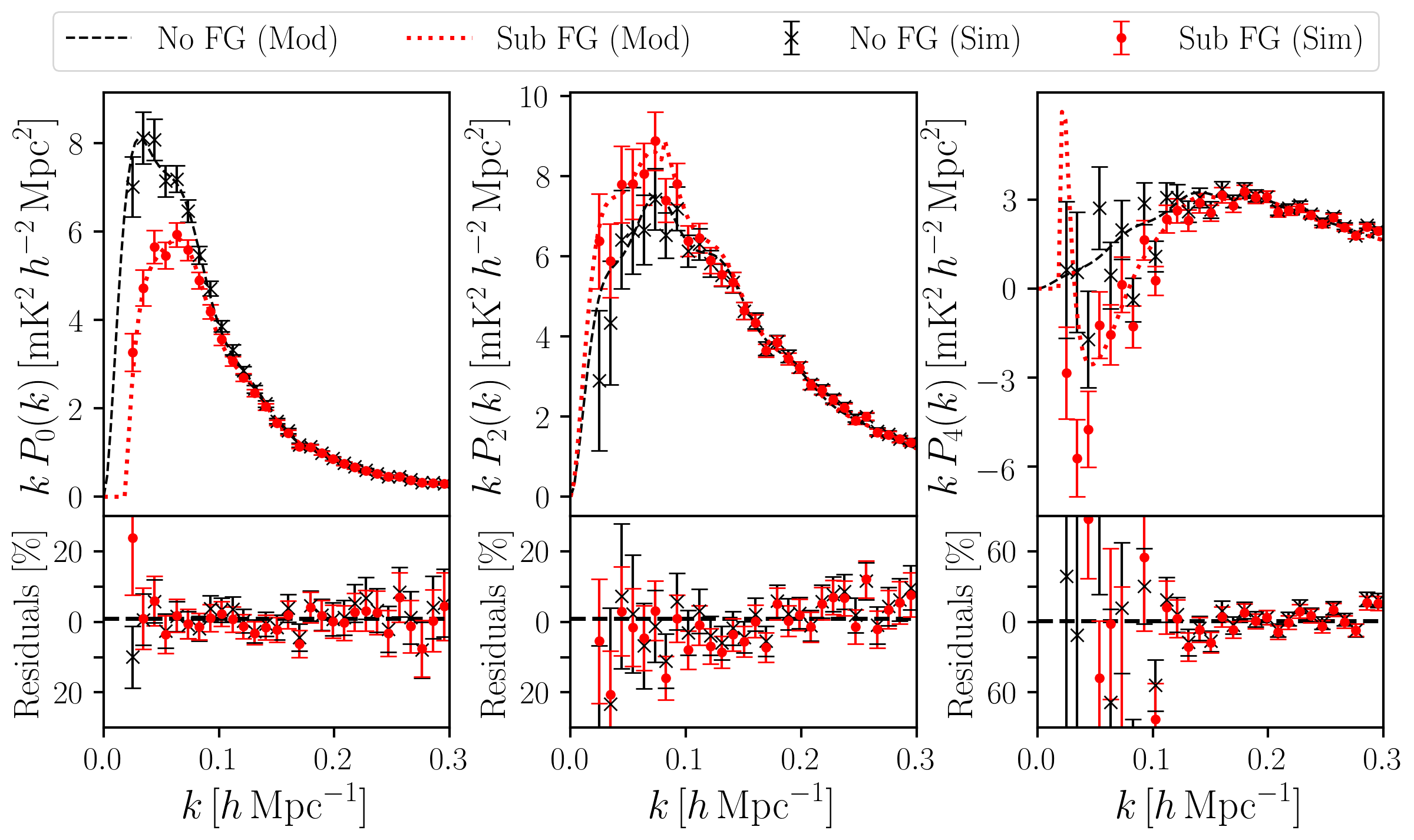}
	\caption{Higher redshift ($z=2.03$) multipole results where a larger beam of $R_\text{beam} = 12\,\text{Mpc}\,h^{-1}$ is used. Instrumental noise is lower and foreground effects are mitigated both due to the larger beam size.}
	\label{MDSAGEMultipolesz2}
\end{figure*}

As discussed, the results in Figure~\ref{MDSAGEMultipolesz0p8} and \ref{MDSAGEMultipolesz0p8wNoise} show some discrepancy between the model (dashed and dotted lines) and simulation. The phenomenological fiducial model we use is demonstrative and not a full fit to the data; therefore, some differences are expected. Whilst in our model we use a HaloFit power spectrum and a FoG factor as approximate non-linear treatments, it is still possible that the discrepancies between the model and simulation could potentially be enhanced at lower redshift where non-linear effects are higher and a more precise treatment is needed. Similar disagreements at low-$z$ have been demonstrated in e.g. \citet{Villaescusa-Navarro:2018vsg} and if these non-linearities are not being modelled effectively, we may find better agreement at higher redshift. 

With this in mind, we ran the simulations at a higher redshift of $z=2.03$ to see if the fiducial model shows better agreement. Figure \ref{MDSAGEMultipolesz2} shows results from the \multidark\ simulated maps at $z=2.03$ and initially shows a better agreement between model and data. However, because this is at a higher redshift, the beam size increases to $R_\text{beam} = 12\,\text{Mpc}\,h^{-1}$ and this is largely why the agreement improves. We tested a smaller beam case and found no significant improvement in comparison to Figures~\ref{MDSAGEMultipolesz0p8} and \ref{MDSAGEMultipolesz0p8wNoise}, suggesting that better modelling prescriptions should be explored when attempting a full analysis for cosmological parameter estimation, especially when we allow for good angular resolution.

For the larger beam results in Figure~\ref{MDSAGEMultipolesz2} we find there is slight improvement in the percentage residuals in the bottom panel, in comparison with Figure~\ref{MDSAGEMultipolesz0p8wNoise}. Whilst we have included instrumental noise in this higher redshift test, the increased beam size means the simulated instrumental noise will be lower (as modelled by equation~\eqref{NoiseEq} keeping the rest of the parameters constant) and this is why we see smaller error bars and better residuals for this higher redshift example.

Figure~\ref{MDSAGEMultipolesz2} is an interesting test using a larger beam and shows that the effects from the foregrounds are much less dominant in this case. This was predicted by our model (see Figure~\ref{ToyFGbigBeam}) and the cause can be summarized as a beam damping of the foreground contaminated modes. We have already discussed this in detail in Section~\ref{BigBeamModelSec}. The distinction between the foreground free (black cross points) and subtracted foreground (red circle points) simulations is therefore less obvious and is only really observed at the smallest-$k$. This is an interesting result since it means intensity mapping experiments with larger beams and realistic levels of noise can be less concerned about foreground contamination when probing anisotropic clustering. However, this assumes that all other systematics are under control and further work using more realistic simulations and pathfinder data is needed. There is also still some disagreement for the largest modes especially for the monopole, thus motivating further understanding of these effects. Furthermore, the quadrupole is still showing evidence of enhancement in the foreground contaminated case and in the hexadecapole, some foreground effects are still evident.

In order to achieve the best agreement for Figure~\ref{MDSAGEMultipolesz2} we found that we needed to use slightly different model parameters for the foreground results. This is expected since at this redshift, the frequency range is lower and therefore the foreground signal will be different compared with the lower redshift results. For the model we use $k^\text{FG}_\parallel=0.019\,h\,\text{Mpc}^{-1}$ and $\mu^\text{FG}=0.13$ for the constant quadrupole cut for modes with $k < 0.08\,h\,\text{Mpc}^{-1}$. Along with these parameters, we also use a bias of $b_\hinospace=1.95$ and find that we achieve a better result if we do not include a FoG factor i.e. we set $\sigma_\text{v}=0$.

\subsection{Parameter Estimation Test}\label{MCMCSec}

To understand the effects of foreground removal on biasing parameter estimation with the \hi power spectrum, we will perform a Bayesian MCMC analysis at redshifts $z=0.8$ and $z=2$, using \hi intensity mapping simulation data and our model. For simplicity we will consider the real space power spectrum (i.e. the monopole without the RSD contribution), to focus on showing the effect of foreground removal on constraining the  amplitude parameter $\Omega_{\hinospace}b_{\hinospace}$.

For this test, in order to allow us to input known parameters that we then attempt to recover, and in order to quickly produce a few realisations of the signal, we simulate intensity maps using log-normal mock catalogs of objects, generated using \texttt{Nbodykit}\footnote{\href{https://nbodykit.readthedocs.io/en/stable//}{https://nbodykit.readthedocs.io}} \citep{Hand:2017pqn}. This method is routinely used for the production of mock galaxy catalogues and is described in \citet{Coles:1991}. For each redshift studied, we obtain 10 different realisations of the \hi signal and average over them.

The mock catalog of objects is generated as follows: First, we specify the central redshift ($z = 0.8$ or $z = 2$), box size, number density of objects and bias of the objects in the simulation. We choose a bias of $1$ for simplicity and a box size corresponding to a fairly large sky area of $40\times40$ deg$^2$ with redshift bin of $\Delta z = 0.4$. At $z = 0.8$ this box has comoving distance dimensions $L_\text{x} = L_\text{y} = 1357\, \text{Mpc}\,h^{-1}$,  $L_\text{z} = 762\, \text{Mpc}\,h^{-1}$ and at $z = 2$ we have $L_\text{x} = L_\text{y} = 2514\, \text{Mpc}\,h^{-1}$,  $L_\text{z} = 400\, \text{Mpc}\,h^{-1}$. Both data cubes have voxel dimensions given by $N_\text{x}=N_\text{y}=N_\text{z}=512$. Using the \textsc{Planck}15 fiducial cosmology \citep{Ade:2015xua}, we define an input linear power spectrum that is used to generate a Gaussian overdensity field in Fourier space, which is then transformed back into configuration space. Next, a log-normal transformation is performed on the overdensity field, which is then Poisson sampled and evolved in time using the Zel'dovich approximation \citep{Zeldovich:1970}. We now have a position for each sampled object, which is converted to a mesh using a mass assignment function (see \citet{Jing:2004fq}). We use the Cloud in Cell (CIC) interpolation with $p=2$.

To provide this log-normal simulation with the observational effects one would expect with an intensity mapping experiment we first multiply the overdensity field by $\overline{T}_\hinospace$, given by equation \eqref{TbarModelEq}. We use a fiducial value of $\Omega_\hinospace = 4.3 \times 10^{-4}$ for both redshifts, consistent with \citet{Masui:2012zc} constraints -- and as already stated we set $b_\hinospace=1$. We then convolve this \hi temperature fluctuation field with a telescope beam. For both redshifts, we use a beam of size $R_\text{beam}=14.41\, \text{Mpc}\,h^{-1}$, which represents a $\theta_\text{FWHM}=1\,$deg beam at $z=0.8$, similar to the beam size of the MeerKAT and SKA-MID dishes.

Finally, we can simulate the effects of foreground removal by adding and subtracting them as described in Section~\ref{FGcontSec}. The final result is a data cube with a \hi temperature fluctuation field including beam and foreground removal effects, where we know the fiducial $\overline{T}_\hinospace$ and $b_\hinospace$. We can then use this simulated map to measure the power spectrum and constrain the amplitude $\Omega_{\hinospace}b_{\hinospace}$, keeping everything else fixed. Since the number density of objects is specified at the start of the simulation, it is straightforward to remove the contribution of shot noise from the calculated power spectrum.

\begin{figure*}
	\centering
  	\includegraphics[width=2\columnwidth]{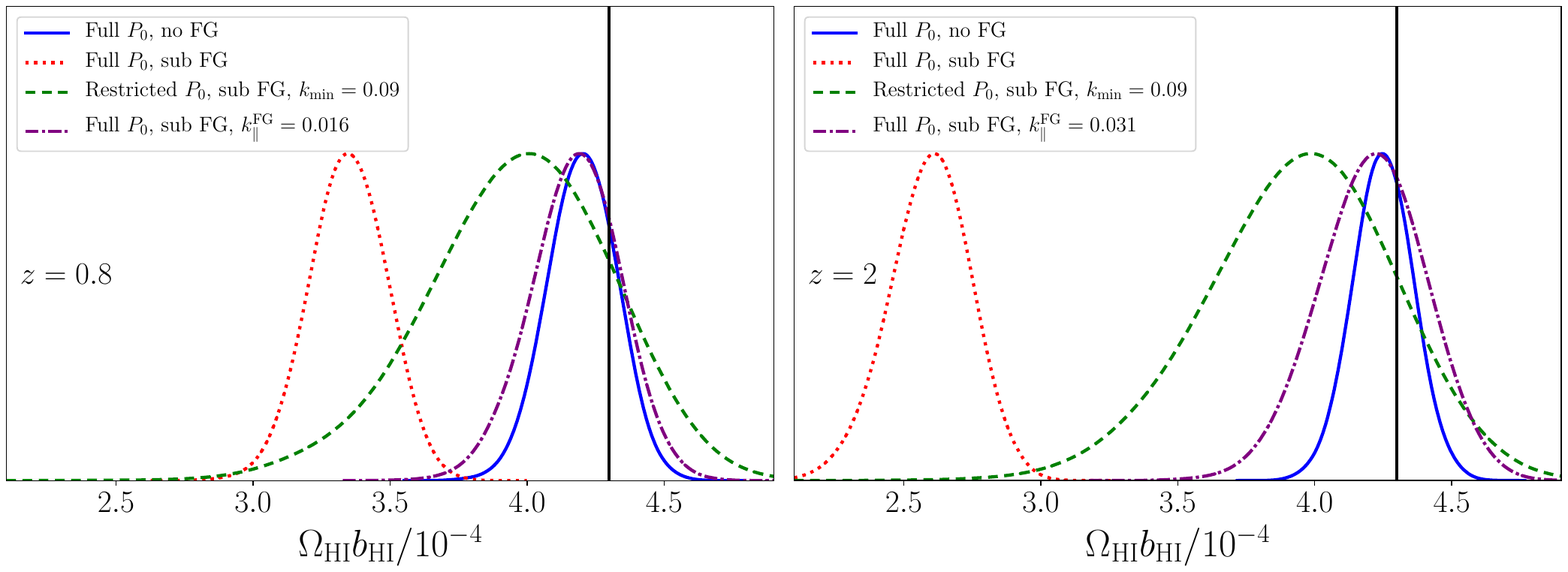}
	\caption{Parameter estimation results for $\Omega_\hinospace b_\hinospace$ and the no RSD $P_0$ model (equation~\ref{ObsMultipole0noRSD}) at $z = 0.8$ (\textit{left}) and $z = 2$ (\textit{right}). The solid blue line represents the case of no foregrounds, dotted red line is the case with  foregrounds added and then removed with \fastica\,  and the dashed green line is the same but with a restricted $k$ range. The dash-dotted purple line is the foreground removed case with a varying $\mu_\mathrm{FG}$ cut. The black solid line is the fiducial (true) value of $\Omega_\hinospace b_\hinospace$.}
	\label{MCMCz08}
\end{figure*}

To model the uncertainties on power spectrum measurements from our simulation, we consider a MeerKAT-like intensity mapping experiment in single-dish mode with a total observation time of 1 week for $z=0.8$ and 10 weeks for $z=2$. The different choices are because we want to have similar error bars for both cases. The noise power spectrum ($P_\text{N}$) due to instrumental (thermal) noise is obtained as in \citet{Pourtsidou:2016imc}, and the uncertainty on the power spectrum is then calculated using equation~(\ref{powerspecerror}).

Given this set of simulated \hi intensity mapping data, and the uncertainties, we wish to test whether the input (fiducial) values of $\overline{T}_\hinospace b_\hinospace$ (equivalently, $\Omega_{\hinospace}b_{\hinospace}$) are recovered using a simple \hi power spectrum model, and how foreground subtraction may bias these results. 
We calculate $P_0 (k)$ following the formalism in Section~\ref{ModellingSec}. Without RSD (and without foreground subtraction), this is given by 
\begin{equation}\label{ObsMultipole0noRSD}
    	P_{0}(k)= \frac{1}{2} \overline{T}_\hinospace^2 b_\hinospace^2 P_\text{M}(k) \, \int_{-1}^{1} \diff \mu\,  \tilde{B}^2_{\mathrm{beam}}(k, \mu)\, .
\end{equation}
We use the above equation in the log-likelihood of our MCMC analysis, varying the parameter $\overline{T}_\hinospace b_\hinospace$, and using a flat positivity prior\footnote{Since this is just a 1-parameter fit an MCMC is not necessary. However we opt for this approach as the accompanying code can be easily extended to include more parameters.}. Note that $\overline{T}_\hinospace$ is dependent on $\Omega_\hinospace$, so by convention, the parameter we are interested in measuring is $\Omega_\hinospace b_\hinospace$. Our fiducial value for this parameter is $\Omega_\hinospace b_\hinospace = 4.3 \times 10^{-4}$. 

The MCMC analysis is performed using the publicly available \texttt{python} package \texttt{emcee}\footnote{\href{https://emcee.readthedocs.io/en/latest/}{https://emcee.readthedocs.io}} \citep{ForemanMackey:2012}, with 200 walkers and 500 samples. For each redshift and beam size we perform four separate analyses; one in the foreground free case and three in the foreground subtracted case. First, we consider the case of no foregrounds. For $z=0.8$, from the survey volume we have $k_{\rm min} = 0.006\, h\,\text{Mpc}^{-1}$ and use bins of width $\Delta k = 0.012\, h\,\text{Mpc}^{-1}$. For $z=2$ we have $k_{\rm min} = 0.005\, h\,\text{Mpc}^{-1}$ and $\Delta k = 0.01\, h\,\text{Mpc}^{-1}$. For both redshifts we impose $k_{\rm max} = 0.3\, h\,\text{Mpc}^{-1}$. We consider the ``full'' $k$-range to be from the survey volume-limited $k_{\rm min}$ to the imposed $k_{\rm max} = 0.3\, h\,\text{Mpc}^{-1}$. When imposing $k$-range cutoffs in subsequent analyses, we refer to the $k$-range as ``restricted''. We now add and remove foregrounds and perform three separate analyses. First, we repeat the analysis in the full $k$-range as in the foreground free case. Next, we restrict the $k$-range considered by imposing a $k_{\rm min}$ value below which the foreground subtraction is damping our monopole signal (as explored earlier in the paper) and thus significantly biasing our parameter estimation results. Finally, we explore the performance of our foreground modelling described in Section \ref{MultipoleFGimpactSec}, using a $k^{\mathrm{FG}}_{\parallel}$ cut for each redshift. We again find that for the monopole a varying $\mu_\mathrm{FG} = k^{\mathrm{FG}}_{\parallel}/k$ works better than a constant $\mu_\mathrm{FG}$ cut. We note that this is for demonstrative purposes only and leave detailed exploration of foreground modelling including RSD and jointly fitting the different multipoles for future work. 

The results for each of these cases can be seen in Figure~\ref{MCMCz08}. It is also common practice in this type of analysis in optical galaxy surveys to limit $k_{\rm max}$ in an attempt to exclude non-linear behaviour 
and obtain less biased results (see e.g. \citet{Markovic:2019sva}). We tried this in our analysis but found no discernible difference in the bias of our results. This is due to the reasonably large beam utilised in our simulated data, which damps the signal and makes non-linearities less significant. If one uses a much smaller beam then this would not be the case anymore and a suitable $k_{\rm max}$ cut should also be imposed.

In both redshifts studied, we find that in the case of no foregrounds, our model for the monopole recovers the input $\Omega_\hinospace b_\hinospace$ within the 1$\sigma$ confidence interval. When we add and subtract foregrounds, we recover a distribution for $\Omega_\hinospace b_\hinospace$ that strongly disagrees with the input, a very biased result in both redshifts. However in both $z=0.8$ and $z=2$ we found that imposing $k_{\rm min}=0.09\,h\,\text{Mpc}^{-1}$ allows us to recover the input $\Omega_\hinospace b_\hinospace$ of our simulation within 1$\sigma$. Note that the width of the distribution (the error) becomes larger when the $k$-range is restricted, as expected. We also find that using our foreground modelling with a varying $\mu_\mathrm{FG}$ cut allows us to recover the input within 1$\sigma$ for both redshifts, by choosing $k^{\mathrm{FG}}_{\parallel}=0.016\,h\,\text{Mpc}^{-1}$ for $z=0.8$ and $k^{\mathrm{FG}}_{\parallel}=0.031\,h\,\text{Mpc}^{-1}$ for $z=2$. The peak of these distributions is closer to the fiducial value than in the $k_{\rm min}$ cut case, and the distributions narrow as expected since we have observed and quantified the model to work well in previous sections. However, the validity of this particular model in the context of real data and more complex foregrounds remains to be studied in future work, while the $k_{\rm min}$ cut method is model-independent. We obtain better results for the no foreground case at $z=2$, which we believe is due to non-linearities being less dominant at this higher redshift. To quantify and summarise the results at each redshift, we note the mean and 1$\sigma$ error of each distribution in Table~\ref{MCMCtable}.

\begin{table}
	\centering
	\begin{tabular}{lcc} 
        \multicolumn{3}{c}{$\Omega_\hinospace b_\hinospace / 10^{-4}$ Estimation} \\
		\hline
		Analysis & $z=0.8$ & $z=2$ \\
        \hline
		Full $P_{0}$, no FG & 4.21 $\pm$ 0.13 & 4.25 $\pm$ 0.11 \\
		Full $P_{0}$, sub FG & 3.35 $\pm$ 0.14 & 2.60 $\pm$ 0.17 \\
		Restricted $P_{0}$, sub FG, $k_{\mathrm{min}}$ cut & 3.98 $\pm$ 0.36 & 3.97 $\pm$ 0.34 \\
		Full $\hat{P}_{0}$, sub FG, $k^\mathrm{FG}_\parallel$ model cut & 4.18 $\pm$ 0.17 & 4.21 $\pm$ 0.20 \\
		\hline
	\end{tabular}
    \caption{Mean and 1$\sigma$ error of $\Omega_\hinospace b_\hinospace / 10^{-4}$ for each posterior distribution obtained with MCMC. For reference, the fiducial (true) value is $\Omega_\hinospace b_\hinospace / 10^{-4}=4.3$.}
    \label{MCMCtable}
\end{table}

The effects of foreground removal dominate below $k_{\rm min} = 0.09\,h\,\text{Mpc}^{-1}$ for both redshifts, but the foreground cleaned case at $z=2$ gives a more biased result compared to $z=0.8$. This suggests that the \textit{scale} at which foreground cleaning contamination begins to bias our results is the same for both redshifts, but the \textit{amplitude} of this bias is larger at $z=2$. This is also demonstrated by the different $k^{\mathrm{FG}}_{\parallel}$ values found to best fit the data with the foreground modelling. The $k^{\mathrm{FG}}_{\parallel}$ value for $z=2$ is almost double that of $z=0.8$, indicating that more signal has been removed alongside the foreground clean at this higher redshift.
Recall that we use beams with the same physical scale for both redshifts, so this effect must be due to other differences between the simulations, namely that we have different box sizes at each redshift. More specifically, the $\Delta z =0.4$ box at $z=0.8$ has a larger radial depth ($L_\text{z} = 762\, \text{Mpc}\,h^{-1}$) compared to the $\Delta z = 0.4$ box at $z=2$ with $L_\text{z} = 400\, \text{Mpc}\,h^{-1}$. This equates to a difference in frequency range with $\Delta \nu = 178\,\text{MHz}$ at $z=0.8$ and $\Delta \nu = 63\,\text{MHz}$ at $z=2$. Probing a wider frequency range makes it easier to identify and remove the foregrounds with \fastica, as more frequency information is available. 

We conclude that in the simple scenario considered, the effect of subtracting foregrounds on parameter estimation is to bias the results significantly. However, we show that by restricting $k$ and effectively discarding the large radial modes most affected by foreground removal techniques, it is possible to retrieve the input $\Omega_\hinospace b_\hinospace$, albeit with larger errors. We also find that it is possible to retrieve this fiducial amplitude parameter using our foreground modelling with a varying $\mu_\mathrm{FG}$ if a suitable $k^{\mathrm{FG}}_{\parallel}$ cut for the data is reasonably known, but leave a more detailed study of the accuracy of this model for future work.

This result is interesting as it demonstrates the effect that the \fastica\ foreground removal has on biasing the $P_0(k)$ power spectrum of the cosmological signal, resulting in biased parameter estimation in the simple scenario we have considered. Similar effects have been seen before in \citet{Wolz:2013wna} using the angular power spectrum $C(\ell)$. Further work is needed in order to determine the best way to deal with the effects of foreground subtraction in order to get unbiased results. Including RSD is essential to accurately model what is actually observed, and further parameters (for example the growth of structure $f$) need to be varied as well -- we plan to do this in future work. This will allow further investigation into the possibility of constraining \hi and cosmological parameters with intensity mapping surveys. More sophisticated theoretical prescriptions might also be required to model the small scales, especially if we consider interferometers like CHIME or HIRAX.

\section{Conclusion}\label{ConclusionSec}

In this paper we have investigated the impact \hi intensity mapping observational effects (telescope beam, foreground removal, and noise) have on anisotropic clustering, and in particular on the \hi power spectrum multipoles. We begun by modelling these effects with a simple input fiducial power spectrum $P_\hinospace(k,\mu)$ from which we constructed the first three non-zero multipoles ($P_0$, $P_2$ and $P_4$), which contain most of the cosmological information. The effects from the telescope beam and foreground cleaning were implemented into the model with the former as a damping function (equation \eqref{BeamDampEq}) and the latter as an exclusion of modes where $\mu < k^\text{FG}_\parallel/k$. This was motivated by the fact that foreground removal is expected to affect the small $k_\parallel$ modes.  Our model is outlined in equation \eqref{ToyFGremovalEq}.
Using simulated intensity mapping data we tested this model and found it can achieve good agreement with the test data.

The simulated data methodology is described in Section~\ref{SimulationsSec}, where we used the \multidark-\textsc{Sage} galaxy data to construct the cosmological \hi signal, while in Section~\ref{MCMCSec} we used log-normal simulations. We also added simulated foregrounds and instrumental noise and performed a \fastica\ reconstruction to produce the foreground cleaned data -- this was then used for comparisons with the foreground free \hi data.\newline
\\
\noindent For clarity, we summarize our main conclusions below:

\begin{itemize}[leftmargin=*]

\item Our model revealed that foreground removal affects each multipole differently. For the monopole ($\hat{P}_0$), we find the foreground clean simply damps the power with a more severe loss for lower-$k$ modes, as  expected. For the quadrupole ($\hat{P}_2$), we find the power is artificially enhanced due to the interaction of the damped power with the Legendre polynomials as explained in detail in Section~\ref{MultipoleFGimpactSec} and demonstrated in Figure~\ref{MultipoleIntegrands}. The hexadecapole ($\hat{P}_4$) also exhibits foreground effects but their behaviour is strongly modulated by the beam size. These findings were also supported by our simulated data tests as shown in Figure~\ref{MDSAGEMultipolesz0p8}.
\\
\item The impact the foreground removal has is modulated by the size of the telescope beam. Figure~\ref{ToyFGbigBeam} shows that increasing the beam renders a larger range of modes (at the small-$k$ end) unaffected by foreground cleaning. This was explained in Sections~\ref{BigBeamModelSec} and is due to the beam damping foreground contaminated modes so that they become a subdominant effect. Again these results were supported by the simulated data measurements that we overlay onto the fiducial model in a similar test shown in Figure~\ref{MDSAGEMultipolesz2}.
\\
\item In Section~\ref{NoRSDSec}, we performed a useful null test by removing the simulated RSD from the data and measure the quadrupole and hexadecapole. Since both of these multipoles should be zero in the absence of RSD, this test reveals the anisotropic observational effects more clearly. Broadly speaking, Figure~\ref{NoRSDMultipole} shows that a foreground clean creates an artificial signal at low-$k$ whilst the beam creates an artificial signal in the higher-$k$ range. These effects thus combine to create a full range of artificial signal across all $k$. Understanding and modelling them properly is needed in order to perform robust RSD measurements with \hi intensity mapping data.
\\
\item The \textit{right} panel of Figure~\ref{kperpkpara} highlighted the main contaminated regions from a foreground cleaning process separated into parallel ($k_\parallel$) and perpendicular ($k_\perp$) components. This revealed that whilst it is largely the small-$k_\parallel$ modes which are most affected, the effect is also $k_\perp$ dependent. The beam effect, on the other hand (\textit{right} panel of Figure~\ref{kperpkpara}), is just a function of $k_\perp$. These results suggest that foreground contamination could be parameterized using the $\mu$ parameter, since the \textit{bottom-right} dark corner is well approximated as a region with $\mu \lesssim 0.2$.
\\
\item Consideration should be given to the above point when constructing an estimator or using the foreground avoidance method for parameter estimation, i.e. how is the region of foreground contaminated modes best defined, as a $k^\text{FG}_\parallel$ cut or a constant $\mu_\text{FG}$ cut. We found that the multipole measurements agree well with our simple fiducial model when using a $\mu_\text{FG} = k^\text{FG}_\parallel / k$ cut, where $k^\text{FG}_\parallel = 0.015\,h\,\text{Mpc}^{-1}$. However, for the quadrupole at $k<0.08\,h\,\text{Mpc}^{-1}$, the model seems to work better with a constant $\mu_\text{FG} = 0.16$ cut. 
These parameters were the ones that returned the lowest percentage residuals in the multipoles (see bottom panels of Figures~\ref{MDSAGEMultipolesz0p8}, \ref{MDSAGEMultipolesz0p8wNoise} and \ref{MDSAGEMultipolesz2}) and the best reduced $\chi^2$ statistics.
\\
\item It is likely that with real data, or with inclusion of more stubborn simulated foregrounds such as polarization leakage, 
more independent components will be needed in the \fastica\ foreground removal.
In that case, the region of foreground contaminated modes becomes larger, but simpler to define. As shown in Figure~\ref{NICcontours}, increasing the number of independent components, $N_\text{IC}$, makes the foreground contaminated region more easily defined with a constant $k^\text{FG}_\parallel$ cut.
\\
\item Our results in Figure~\ref{NICcontours} also suggest that arbitrarily going to high levels of $N_\text{IC}$ does not necessarily cause results to converge.
\\
\item Lastly we showed that using our model or a restricted $k$-range we were able to recover the correct $\Omega_\hinospace b_\hinospace$ value within $1\sigma$ (see Figure~\ref{MCMCz08}). Without accounting for the effects of foreground removal, the results were very biased. This highlights the importance of understanding the extent of foreground contamination when working with real data.
\\
\item The accompanying \texttt{python} toolkit is available at \url{https://github.com/IntensityTools/MultipoleExpansion}, and includes maps and the relevant data required to reproduce our results. We also include a set of pedagogical Jupyter notebooks.

\end{itemize} 

Our simulations have omitted complications from polarization leakage of foreground synchrotron radiation, which can disrupt the frequency coherence of the foreground signal rendering them more challenging to remove. Although there is little work on the effects of polarization leakage on low-redshift intensity maps, the hope is that a more aggressive foreground clean, coupled with precise instrument calibration, should mitigate the effect of this. Whilst we included realistic levels of instrumental noise, this was uncorrelated (white) noise. We have ignored the effects of frequency correlated ($1/f$) noise, with the assumption that they can be mitigated with appropriate scanning strategy and calibration. Further treatment of these effects  would be worth pursuing.

In future work we plan to investigate a suite of different models and perform fitting and cosmological parameter estimation analyses using MCMC, extending previous works \citep{Pourtsidou:2016dzn, Bacon:2018dui, Castorina:2019zho, Bernal:2019jdo}. 
This will also look into more detail of the optimal approach for dealing with the foreground contaminated regions. Another aim would be to include  cross-correlations with simulated optical galaxy data.

Using the multipole expansion formalism with redshift space models including observational effects is the standard way to constrain cosmological parameters with data from large scale structure surveys. Doing the same with \hi intensity mapping data is necessary to fully exploit the constraining power of large sky radio cosmological surveys, and enable cross-correlation analyses across a wide redshift range. The results in this work demonstrate that, whilst there are several challenges, these should be surmountable with the correct understanding of the observational effects. We aim to use the simulations, models, and numerical tools developed in this paper to help analyse \hi intensity mapping data from MeerKAT very soon.\newline
\\
 
\section*{Acknowledgements}
We are grateful to Tzu-Ching Chang, Kiyoshi Masui, Laura Wolz, Julian Bautista, and Marta Spinelli for very useful discussions.
SC is supported by STFC grant ST/S000437/1, and also acknowledges support by the University of Portsmouth for part of this project. AP is a UK Research and Innovation Future Leaders Fellow, grant MR/S016066/1, and also acknowledges support by STFC grant ST/S000437/1. PS is supported by the Science and Technology Facilities Council [grant number ST/P006760/1] through the DISCnet Centre for Doctoral Training. DB is supported by STFC grant ST/N000668/1. This research utilised Queen Mary's Apocrita HPC facility, supported by QMUL Research-IT \url{http://doi.org/10.5281/zenodo.438045}. We acknowledge the use of open source software \citep{scipy:2001,Hunter:2007,  mckinney-proc-scipy-2010, numpy:2011,  Lewis:1999bs,Lewis:2019xzd}. Some of the results in this paper have been derived using the \texttt{healpy} and \texttt{HEALPix} package. We thank New Mexico State University (USA) and Instituto de Astrofisica de Andalucia CSIC (Spain) for hosting the Skies \& Universes site for cosmological simulation products.

\bibliographystyle{mnras}
\bibliography{Bib} 


\bsp	
\label{lastpage}
\end{document}